# Patterned states in the nematic phase of flexible-core phenyl benzoate dimers

K. S. Krishnamurthy,[1*] Santosh Y. Khatavi,[2] Channabasaveshwar V. Yelamaggad,[1] and N. V. Madhusudana[3†]

[1]*Centre for Nano and Soft Matter Sciences, Survey No. 7, Shivanapura, Bangalore 562162, India*
[2]*National Forensic Science University, Dharwad 580011, India*
[3]*Raman research Institute, Bangalore 560080, India*

ABSTRACT

This study deals with both spontaneously-formed and electrically-induced structures observed in the nematic phase of two dielectrically negative twist-bend nematogens (dimeric phenyl benzoates). In planar cells, the nematic layers are inhomogeneous, existing in a quasiperiodic *ground* state that involves essentially azimuthal director deviations. This phenomenon is understood using a simple model based on the relative flexoelectric and elastic contributions to free-energy. In an external electric field, depending on the conditions of excitation, a variety of instabilities are obtained. In an increasing static field, for example, the initial periodic *surface* electroconvective instability is followed by the Bobylev-Pikin *volume* flexoelectric instability. Uncommonly, the growth of the latter takes place via nucleation and front propagation. As in rigid bent-core systems, due to the low bend elastic deformation cost, flexoelectric bands progressively distort as they narrow (mediated by edge dislocations) under an increasing field to eventually form fanlike objects. In the mHz region, in each half cycle following 0 V, the electroconvective and flexoelectric instabilities appear transiently with the latter setting in at a higher voltage. Above a few Hz, flexoelectric instability ceases; the sequence of patterned states then is oblique roll→bimodal grid→normal roll→chevron. In the very high frequency regime (>100 kHz), periodic wide bands oriented normal to the rubbing axis are obtained at a threshold voltage that decreases with increasing frequency, just as in the case of high frequency nonstandard prewavy instability in rigid bent-core systems. Our measurement of threshold voltage corresponding to different frequencies (or electrical conductivity and permittivity values) in this regime agrees well with the earlier theoretical predictions relating to the inertial conduction instability.

---

*Contact author: murthyksk@cens.res.in
†Contact author: nvmadhu@rri.res.in

## I. INTRODUCTION

Dimers of achiral polar molecules with an odd-numbered methylene linkage are best known for their two uniaxial nematic (N) phases. The conventional N phase, in which the molecules prefer to be along a unit vector (director) **n**≡–**n**, upon cooling, undergoes a local-reflection-symmetry-breaking, first order transition into the twist-bend nematic ($N_{TB}$) phase at a temperature $T_{TB}$. In the latter, also described as the heliconical phase, the molecular director **m** precesses conically around a macroscopic uniaxial direction, or twist director **t,** with a nanometric pitch. The $N_{TB}$ phase visualized very early in the context of spontaneous flexoelectric (flexo-, for short) polarization [1], and predicted more recently to occur under conditions of negative bend elastic modulus $k_{33}$ [2], was experimentally realized later in dimers such as 1",7"-bis(4-cyanobiphenyl-4′-yl)heptane (CB7CB) [3] and CB11CB [4], and conclusively identified [3]. The $N_{TB}$ phase, by virtue of its many distinctive attributes—like the spontaneously-formed striped ground state [4-6], nanoscale modulation [7-10], remarkably distinctive electric [11-19] and magnetic [5] responses—has remained a subject of active theoretical and experimental interest [20-23].

The nematic phase made of bent molecules, regardless of their core being rigid or flexible, has also attracted much attention. One of the reasons for this has been the extraordinary elastic behaviour found in the N phase of bent-core systems in general, and soft-bent dimers in particular [3, 5, 24-27]. In a significant study of various physical properties of nematic CB7CB [27], the bend elastic modulus $k_{33}$ from capacitance measurements is found, while cooling, to decrease to a minimum of 0.38 pN before showing a pretransitional rise toward $T_{TB}$. In most electric field experiments conducted in the N phase of twist-bend nematogens, the compounds used are dielectrically positive, being of the type CBnCB. For example, CB11CB [28] and CB7CB [29] have been used in flexoelectrooptic experiments that show the effective flexocoefficients to be considerably larger than in calamitics. Similarly, electro-optic measurements in CB7CB confined to HAN cells indicate a very large bend flexo coefficient $e_3$ [30]. Under electrical excitation, nematic CB7CB is observed to exhibit confined electroconvective and wormlike flexoelectric instabilities [31]. In CB7CB doped with a surfactant, twisted bipolar N drops are found to show the Lehmann rotation with a velocity that follows a scaling law involving the temperature gradient, surface twist angle and drop radius [32]. In a recent study on the stability of different Freedericksz states in elastically varying compounds held in pi-cells, nematic CB7CB is examined comparatively with both rigid bent-core and calamitic compounds [33].

The present work focuses on both spontaneously-obtained and electrically-induced patterned states in the *nematic* phase of two twist-bend nematogens, which are, unlike CB7CB, negative in dielectric anisotropy. The primary motivation for the study was to compare the electric instabilities obtained with those of rigid bent-core materials, particularly in regard to the so-called nonstandard prewavy (PW) states of the latter [34]. As it turned out, in addition to the electric instabilities, we came across a striped state in the absence of any external field. One of the objectives here is to offer a simple model for its physical origin. Among other results, we also observe the very high-frequency PW-like instability and find it to conform to the inertia mode [35]. These results are presented and discussed in Sec. III that follows the experimental section below.

## II. EXPERIMENTS

The nonamethylene-linked dimers of phenyl 4-pentyloxy and hexyloxy benzoates (Fig. 1)—P5OBD9 (5O9, for short) and P6OBD9 (6O9, for short)—which were first reported by Mandle along with several other twist-bend nematogenic compounds [36]—were synthesized by two of the authors (SYK and CVY). As determined by polarization microscopy, the phase sequence in

R=$C_5H_{11}$ in P5OBD9 and $C_6H_{13}$ in P6OBD9

FIG. 1. Twist-bend nematogens examined are dimers of phenyl 4-pentyloxy and hexyloxy benzoates with a nonamethylene linkage, abbreviated as P5OBD9 and P6OBD9 (more briefly, 5O9 and 6O9), respectively.

5O9 was Isitropic, I (88.2 ºC) Nematic, N (73.2 ºC) twist-bend nematic, $N_{TB}$; and that in 6O9 was I (94.5 ºC) N (81.5 ºC) $N_{TB}$ (74.4 ºC) Smectic $C_A$, $SmC_A$ (~46.5 ºC) crystal, Cr. The $N_{TB}$ phase in 5O9 supercools and transforms into the crystal at about 46 ºC. The identification of $SmC_A$ phase in 6O9 is from [36]. For optical study, we used a Carl-Zeiss Axio Imager.M1m polarizing microscope with an attached AxioCam MRc5 digital camera; the AxioVision software enabled recording time-lapse and *z*-stacked images. Transmitted light measurements were done using a photodiode (Hamamatsu S2281) connected to a wideband amplifier (Hamamatsu C9329); intensity profiles were recorded on a PicoScope (Model 4292). For imaging with monochromatic light in the transmission mode, a He-Ne laser scanning confocal attachment (Carl Zeiss LSM5) was employed. Samples were held in commercial sandwich cells (from AWAT, Poland and Instec, USA), made of ITO glass plates, coated with polyimide layers; planar alignment was ensured by unidirectional rubbing (along $R$) such that the pretilt angle remained within 3º; we used both parallel and antiparallel rubbed cells. The sample temperature $T$ was held constant to an accuracy of ±0.1˚C using an Instec HCS402 hot-stage along with a STC200 controller. Relative temperature $T_r$ in the text stands for $T–T_{TB}$, where $T_{TB}$ is the onset temperature of the $N_{TB}$ phase. A Stanford Research Systems function generator (DS345) coupled to a FLC Electronics voltage amplifier (model A800) was used in electric field studies. The frequency $f$ and voltage $U$(rms) of the applied field were measured with a Keithley-2002 multimeter. Fig. 2 shows the experimental arrangement. We denote by P($\alpha$)–A($\beta$) the setting of the polarizer P and analyser A with their axes at angles $\alpha$ and $\beta$ (degrees) relative to $x$.

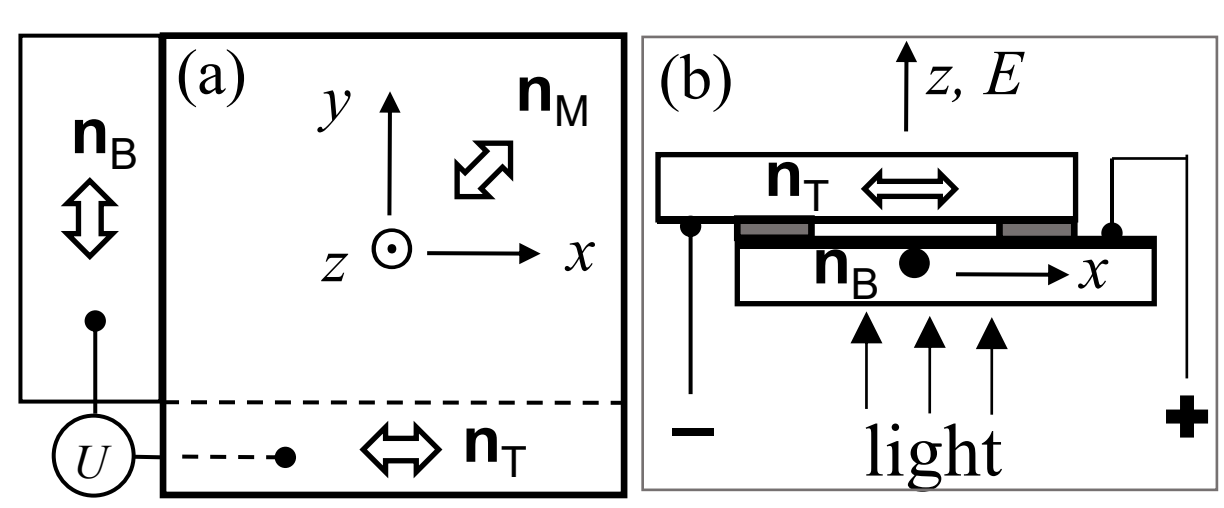

FIG. 2. Schematic of the experimental geometry for a 90°-twist cell; (a) top-view, and (b) side-view along *y*. $\mathbf{n}_T$ and $\mathbf{n}_B$ denote the nematic directors at the top and bottom substrates respectively; and $\mathbf{n}_M$ denotes the midplane director. The field is taken positive when acting along *z*, as in (b). For an untwisted planar cell, the alignment in the nematic layer is uniformly along *x*.

## III. RESULTS AND DISCUSSION

The modulated states of the director field found in the nematic phase of the two phenyl benzoates may be considered as arising from one or the other of the following phenomena: (a) spontaneous quasiperiodic distortion defining the ground state, (b) electroconvection (EC) localized near the electrodes and arising in static and low-frequency fields, (c) Bobylev-Pikin volume flexoelectric (flexo-, for short) distortion arising in static and very low-frequency fields, (d) EC originating in bulk in the low-frequency (5 Hz-10 kHz) regime, (e) Very high frequency (MHz) inertia mode of EC. Each of these states is described and discussed in the succeeding subsections.

### A. The striped ground state (SGS)

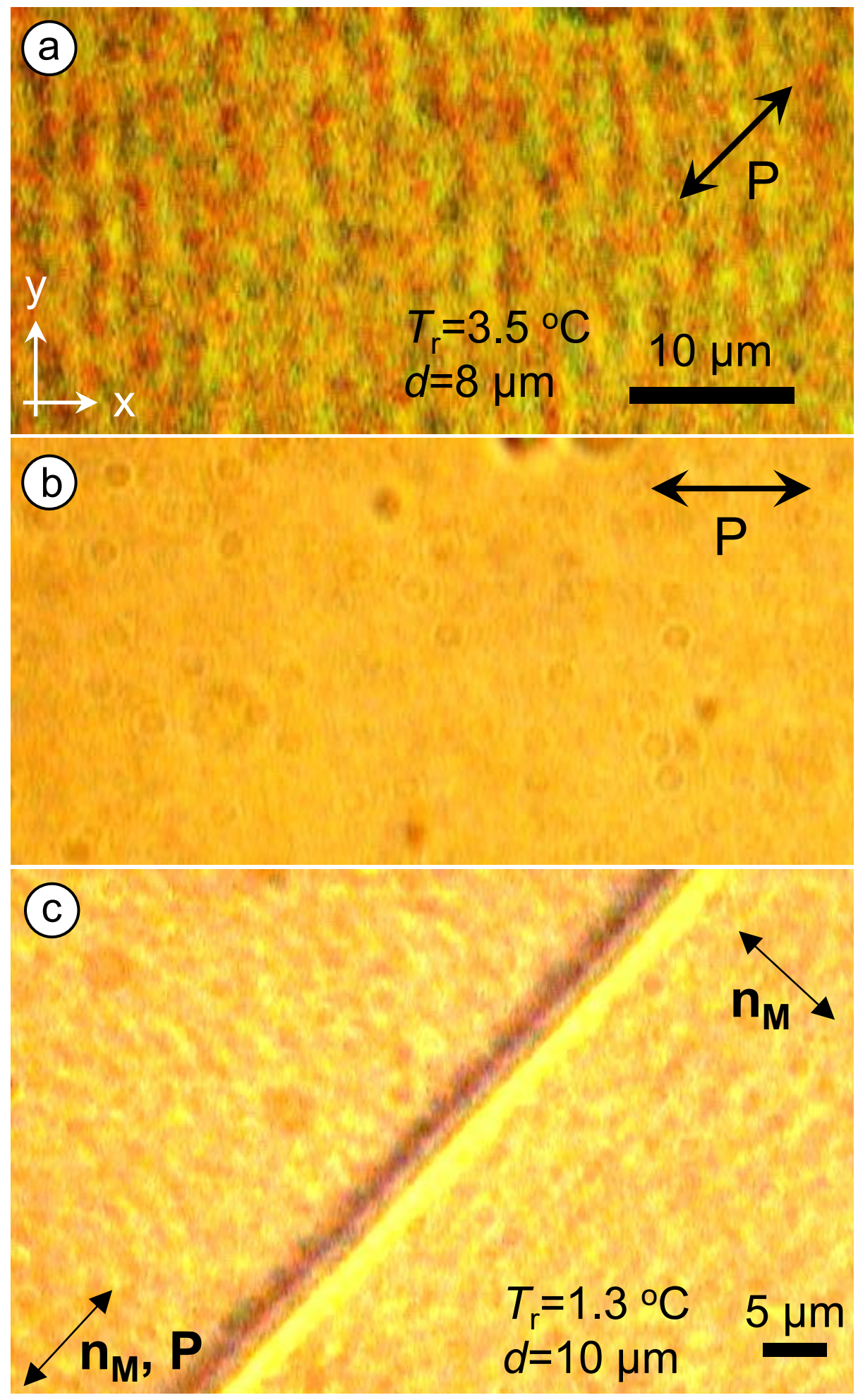

FIG. 3. (a) Inhomogeneous patterned appearance of a nematic layer of 6O9 in a planar cell viewed with a single polarizer set diagonally; maximum visibility is obtained for P(45) or P(135). (b) Uniform appearance of the layer for a polarizer with its axis along the rubbing direction R||$x$ (or along $y$). (c) In a 90° twisted planar cell, the stripes appear normal the midplane director $\mathbf{n}_M$ (disposition of $\mathbf{n}_M$ is revealed by flexoelectric domains formed in a static field); here, the twist is right-handed below the diagonal disclination and left-handed above it.

Both the compounds, 5O9 and 6O9, examined in the sandwich geometry, in cells treated for planar alignment through unidirectional buffing, behave identically in terms of their ground, field-free state. They present an inhomogeneous appearance with quasiperiodically ordered short stripes stretching along the normal to the rubbing axis ($R_\perp||y$) and defined by a wave vector $\mathbf{q}_x$ broadly along the rubbing axis $R||x$. The separation between adjacent stripes is a few μm; it is not affected predictably by the sample thickness. The pattern contrast, which is generally very low, increases progressively as the $N_{TB}$ transition is approached while cooling the nematic from about 10 ºC above $T_{TB}$. The visibility of stripes is dependent on the setting of polarizer(s); notably, it is maximum for a single polarizer, P or A, set at 45° or 135° to the rubbing axis [Fig. 3(a)]; by comparison, for a polarizer set at 0° or 90° to $x$, the contrast practically vanishes [Fig. 3(b)]. Similarly, between diagonally crossed polarizers, the inhomogeneity is masked by the general brightness. However, with a marginal uncrossing of polarizers along *x* and *y*, as for example with P(10)–A(90), the periodic structure becomes discernible. Significantly, the pattern is *dynamic* with the intensity at a given location fluctuating randomly between minimum and maximum at an interval of a fraction of a second. We also observe a similar intensity fluctuation in the ground state stripes pattern in planar samples of the well-known twit-bend nematogen CB7CB. In order to find whether the nonuniform texture is a surface or bulk feature, we examined planar layers held in quarter turn twist cells. In such samples, the stripes formed diagonally, in the direction orthogonal to the midplane director $\mathbf{n}_M$ [Fig. 3(c)]; the orientation of $\mathbf{n}_M$, as described later, is ascertained from flexoelectric domains that form parallel to it under excitation with dc fields. The SGS in unaligned or degenerate tilted planar samples provides a straight forward means of identifying the director field around line disclinations formed along the layer normal. This is illustrated in Fig. 4, wherein some line defects with strength s=1 and ±1/2 are present. Since the stripes locally represent the director normal $\mathbf{n}_\perp$, the **n**-field around +1 defect consists of concentric circles. In general, the director and director-normal fields around any linear defect are represented by a pair of orthogonal curves.

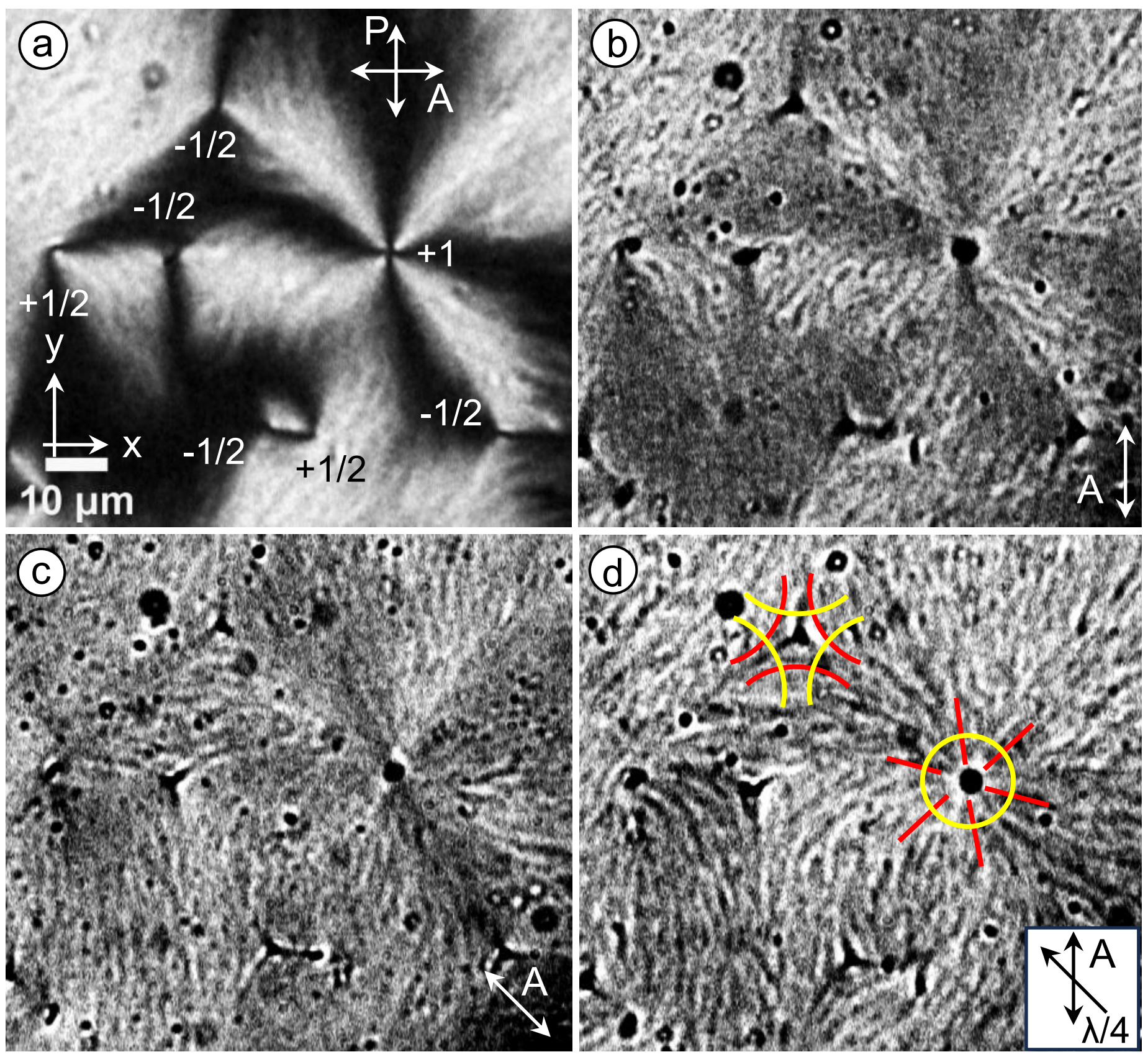

FIG. 4. Gray scale images of the schlieren texture in a degenerate tilted planar 6O9 sample at $T_r$=3.5 °C, for different polarization contrasts. With crossed polarizers, P(0)–A(90), line defects of different strengths are discernible from the feeble ground state stripes. With only the analyzer along *y,* A(90), diagonal stripes appear with maximum contrast (b). With a diagonal analyzer A(135), stripes along *x* and *y* are best defined (c). Addition of a quarter wave plate to A(90) renders the stripes visible all over (d). At any given location, the stripe lies along the director normal $\mathbf{n}_\perp$; by way of example, lines of **n** and $\mathbf{n}_\perp$ fields are shown in yellow and red, respectively, around two defects in (d).

### B. A simple model for the patterned ground state

The spontaneously obtained striped pattern just described is similar to the ground state modulation previously observed in other bent-core systems studied previously in the *unaligned* schlieren geometry [37]; it was then interpreted as a surface phenomenon. Now that the modulation has come to be recognized as a bulk phenomenon, its explanation calls for a different approach; in what follows, we offer a simple model to explain the SGS, which takes into account the relative flexoelectric and elastic contributions to free-energy.

The notable feature brought out in Fig. 3 is that, in the higher temperature nematic phase, a planar aligned sample exhibits *fluctuating* quasiperiodic structures with the wave vector **q** along the rubbing direction. While their optical visibility is generally low, with a polarizer diagonal to **q** they are clearly discerned. We may take the director (**n**) field in the SGS as essentially comprising azimuthal distortions, in planes parallel to the cell plates (Fig. 2). The observation in 90$^{o}$-twisted planar samples [Fig. 3(c)] also demonstrates that the maximum deviation of **n** occurs at the midplane of the sample.

The curvature elasticity of nematic liquid crystals is weak, resulting in strong thermal fluctuations of **n** [38] which, for example, leads to a strong scattering of light.  Bent shape of molecules can be expected to reduce the bend elastic constant $k_{33}$ and measurements on several compounds have shown that the higher temperature nematics made of such molecules are characterised by $k_{11}>k_{22}>k_{33}$ [27], where $k_{11}$ and $k_{22}$ are the splay and twist elastic constants, respectively. Indeed, the director distribution around defects in the compounds used in the present study are dominated by bend distortion (Fig. 4). We can expect that the thermal fluctuations of the director are dominated by bend distortion in these compounds. In a planar aligned sample with strong anchoring of **n** at the boundaries, the fluctuation amplitude can be expected to be highest at the midplane of the cell. Fluctuations in the *xz* plane, which is orthogonal to the bounding surfaces, involve the tilt angle $\theta$ of **n**, and to satisfy the strong surface anchoring, the relevant distortion will be a combination of bend and splay of **n**. On the other hand, fluctuations in the *xy* plane parallel to the boundaries involve the azimuthal angle $\varphi$ of **n**, the relevant distortion being a combination of bend and twist of **n**. As $k_{11}$ is much larger than $k_{22}$, the fluctuations are dominated by the azimuthal distortion of **n**.

The quasi-periodic structures observed in the fluctuations can be described by the following relation:

$$\varphi(x,z) = \varphi_0 \cos\frac{\pi z}{d}\sin(qx) \qquad (1)$$

in which $\varphi_0$ is the maximum azimuthal angle at the mid-plane with $z$=0, and $d$ is the sample thickness, with the bounding surfaces at -$d$/2 and +$d$/2. The local quasiperiodic structure extending over a few spatial periods along the $x$-axis is described by the sine function. The structure also extends along the $y$-axis over a length $\xi$ comparable to the sample thickness, and we assume that the distorted structure has no $y$-dependence over this length. Our main interest is to gain a physical understanding of the quasi-periodic order rather than the more typical nonperiodic fluctuations in the sample. Near the threshold, the amplitude $\varphi_0$ is small, and we limit the calculations of the relevant energies only up to quadratic terms in $\varphi_0$.

The elastic energy density is given by

$$F_{elast}(x,z) = \frac{k_{33}}{2}\left(q^2 cos^2\frac{\pi z}{d}cos^2 qx\right)\varphi_0^2 + \frac{k_{22}}{2}\left(\frac{\pi^2}{d^2}sin^2\frac{\pi z}{d}sin^2 qx\right)\varphi_0^2 \qquad (2)$$

The periodic bend distortion of **n** generates a periodic flexoelectric polarization in the medium, and several measurements [28-30] have shown that the relevant coefficient $e_3$ is likely to be rather large. Again keeping only the leading term in $\varphi_0$,

$$\boldsymbol{P}_{flex} = \boldsymbol{j}\, e_3\, q\, \varphi_0 \cos\frac{\pi z}{d}\cos qx \qquad (3)$$

in which $\boldsymbol{j}$ is the unit vector along the $y$-axis. Neighbouring half waves have opposite orientations of flexoelectric polarization, and their mutual geometrical disposition results in a negative interaction energy between them (Fig. 5).

As the fluctuating periodic structures are relatively long lived, lasting for ~ 1 second, we can expect that their net energy will be close to 0. We simplify the analysis by noting that the maximum polarization is located at the mid-plane of the cell ($z$=0), and for $qx$=0 or an integral multiple of $\pi$. We integrate the flexoelectric polarization given by Eq.3 over the sample thickness $d$ and half the wavelength ($\pi/q$) centred around $x$=nπ. We also make the simplifying assumption that the resulting dipole moment per unit length $\boldsymbol{p}$ acts along lines along the $y$-axis at $z$=0 and $x$= nπ, and is given by

$$\boldsymbol{p} = \pm\boldsymbol{j}\frac{4d}{\pi}e_3\varphi_0 \tag{4}$$

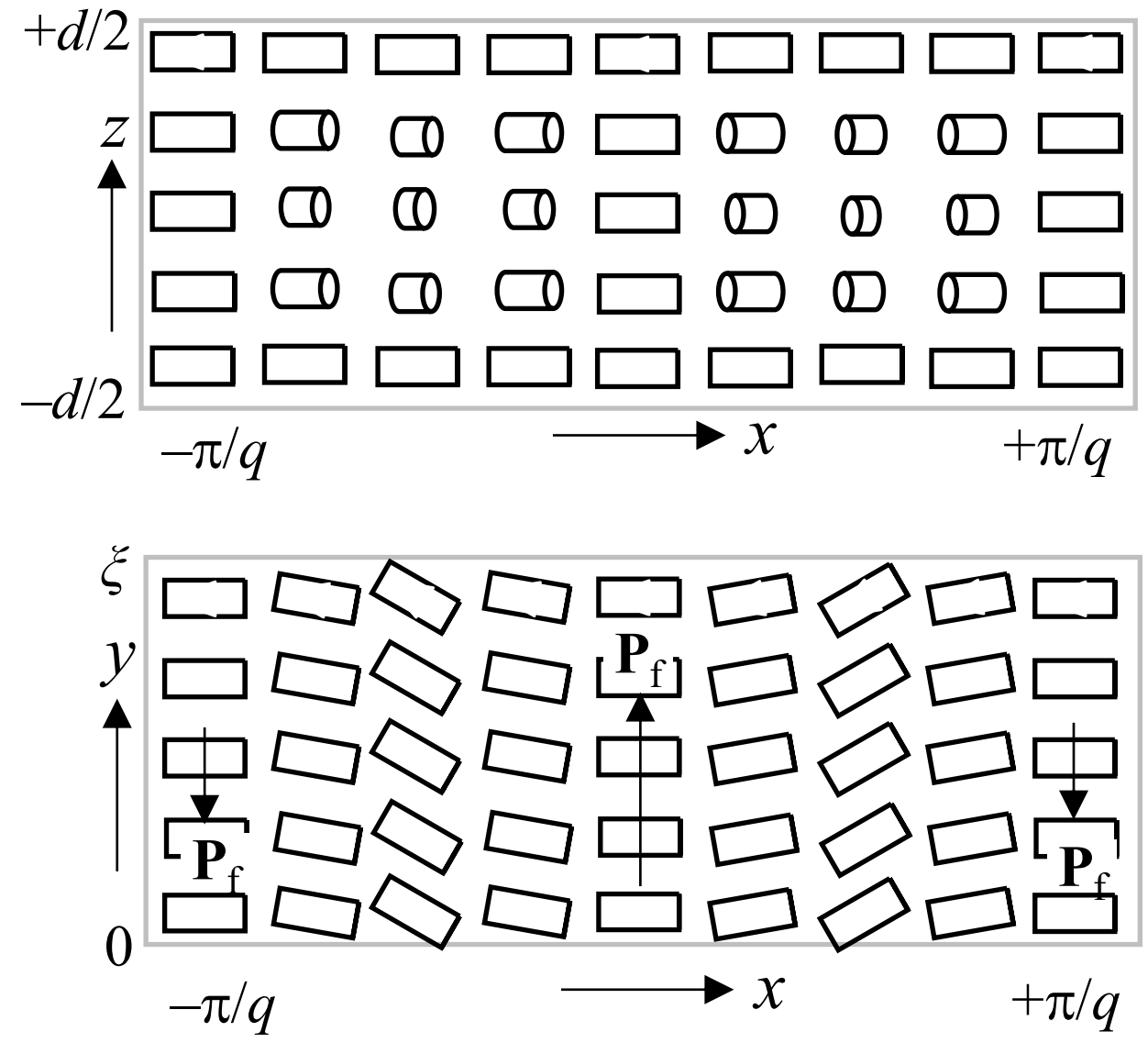

FIG. 5. Schematic of the spontaneously obtained periodic deviations of the director **n** that result in the striped ground state. While, in the layer plane *xy*, the distortion is entirely of the azimuthal type, that in the plane *xz*, involves the twist. $\mathbf{P}_f$, $q$ and $\xi$ represent flexo-polarization, wave vector and average stripe length along $y$, respectively.

The +(–) sign in the above equation corresponds to n being even (odd).

Experimentally, the fluctuation extends over a length $\xi$ (~$d$) along the $y$-axis. The bounding surfaces in the $xz$ plane are covered by polarization charges which create locally an electric field opposed to **P**. However, as the surface charge densities of neighbouring half waves have opposite signs, which are also screened by free ions in the medium, the surface charge contribution to the energetics can be ignored. The dipolar interaction energy between neighbouring half waves can be estimated as

$$E_{flex} \approx -\frac{1}{4\pi\varepsilon_0}\frac{(4de_3\xi/\pi)^2}{(\pi/q)^3}\varphi_0^2 \tag{5}$$

in which $\varepsilon_o$ is the vacuum dielectric constant. We may note that the actual energy will be lower than the above in a more detailed calculation. Though there are two neighbours for each half wave, as each is shared between two neighbours, Eq. (5) is the energy of each half wave, to be compared with the positive elastic energy needed to create the relevant distortion. We also ignore the positive dipolar energy between second neighbouring half waves which is 8 times smaller than that given by Eq. (5).

The total elastic energy cost of the director distortion in half a wave is calculated by integrating $F_{elast}$ given by Eq. (2) over the relevant volume, and is given by

$$E_{elast} = \frac{\pi\xi}{8}\left(k_{33}dq + k_{22}\frac{\pi^2}{dq}\right)\varphi_0^2 \tag{6}$$

The net energy of the fluctuation is given by

$$E_{fluct} = E_{flex} + E_{elast} \tag{7}$$

This has to be close to zero for the fluctuation to live long enough. From the experiment (Fig. 3), $\xi \sim d$, and we assume $\xi = d$. Experimental measurements on dimers which exhibit the $N_{TB}$ phase show that $k_{33}$ is about 1 pN and $k_{22}$ is about 1.5 pN a few degrees above the transition to the $N_{TB}$ phase [27], while $e_3$ is about 10 pC/m$^2$. Using these inputs, the only free parameter to decide the sign of $E_{fluct}$ is $dq$. It crosses 0 when $dq \approx 10$, or the wavelength $\lambda \approx 0.63d$, which is similar to the observed value. The energy becomes more negative as $dq$ is increased, i.e., as the wavelength is decreased, and the analysis has to be extended to include the $\varphi_o^4$ term to find the equilibrium value. But our simplified calculations clearly show that the low value of $k_{33}$ coupled with an enhanced magnitude of $e_3$, both of which arise from the bent geometry of the dimers, give rise to the periodic fluctuations involving only the azimuthal angle $\varphi$ of the director in the high temperature nematic phase of the compounds studied.

The sample undergoes a transition to the $N_{TB}$ phase when it is cooled to the relevant transition temperature. An elastic model has been proposed to show that a negative value of $k_{33}$ results in this phase [2]. It has been argued that a coupling of bend distortion and $e_3$ which is assumed to take a divergent value as temperature is lowered can lead to the required negative sign of an effective $k_{33}$. A typical value of the pitch is only about 10 nm near the transition point, decreasing further at lower temperatures. This is only 2 or 3 times the molecular length and a natural question is whether director elasticity plays a decisive role at this length scale. Recent detailed X-ray scattering studies on the $N_{TB}$ phase [23] have shown that the bent molecules form a double helical structure, aided by the biphilicity of the molecules. In other words, short range molecular interactions of the bent molecules give rise to the $N_{TB}$ structure. On the other hand, as we have argued, the periodic fluctuations seen in the higher temperature nematic phase result from the low bend elastic constant and an enhanced flexoelectric coefficient of the dimers with bent geometry.

### C. Surface electroconvective and volume flexoelectric states excited by static fields

In an untwisted planar sample of either dimer, the effect of applying a voltage $U$ is dependent on whether this voltage is reached gradually, in small steps, or applied all-at-once. In the former case, as $U$ increases, while no definitive periodic pattern is obtained, the amplitude of director fluctuations continues to rise until the voltage equals the flexoelectric threshold $U_f$ [compare Figs. 6(a) and 6(b)]. At $U_f$, periodic domains with well-defined wavevector $\mathbf{q}_y$ (transverse to the alignment direction $\mathbf{n}_o$) nucleate randomly in space and time, and begin to grow [Fig. 6(c)]. By contrast, when a large enough voltage (above 4.2 V at $T_r$=2 ºC in 5O9) is suddenly applied, there appears soon after a pattern of zig-zag stripes with the wave vector close to $x$ and period, about the sample thickness $d$ [Fig. 6(d)]. This pattern is transient, decaying toward the base state over a few seconds [Fig. 6(e)]. Additionally, when using a single polarizer P, at threshold, it is best seen for P(0) (i.e., when the electric vector of light is along the rubbing direction) and practically absent for P(90). These features of the zig zag pattern correspond to a periodic modulation that is primarily of the splay-bend type at threshold, belonging to the electroconvective instability. At $U$=$U_f$, as before, domains parallel to $\mathbf{n}_o$ nucleate slowly and grow in what appears a patternless background left by the initial zig-zag domains [Fig. 6(f)]. That these domains are due to flexoelectric distortion is clear from their distinguishing features. For example, they are observed in natural light as periodic bright focal lines produced by the extraordinary component of light passing a periodically tilted structure, with the tilt-angle expressible as $\theta = \theta_o \cos(qy) \cos(\pi z/d)$, with $q = 2\pi/\lambda_F = \pi/w_F$, where $\lambda_F$ is the period of modulation and $w_F$ is the domain width; the domains also exhibit vivid interference colors under polarization contrast due to the added periodic azimuthal distortion $\varphi = \varphi_o \sin(qy) \cos(\pi z/d)$. This $(\theta, \varphi)$ modulation represents the well-known Bobylev-Pikin (BP) flexoelectric volume instability observable optically regardless of the state of polarization of incident or transmitted light. However, unlike the usual BP instability in rodlike nematics that sets in *extensively* at $U_f$ as a supercritical bifurcation, the transformation here is localized, nucleating randomly in space and time, and developing outward through front propagation. Figure 7(a) depicts the typical appearance of growing flexoelectric islands in 5O9;

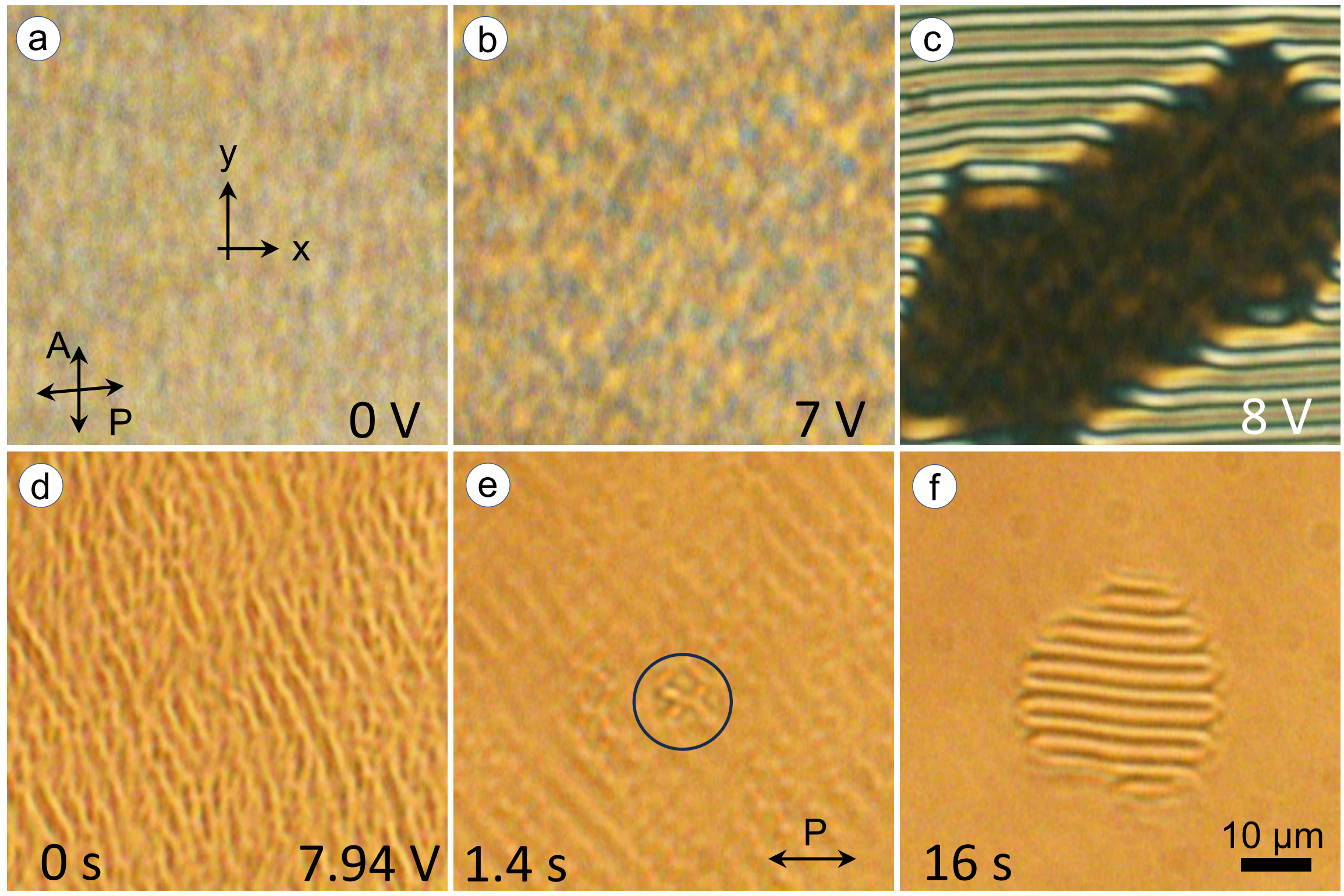

FIG. 6. The effects obtained in a static field vary depending on whether a chosen voltage is applied suddenly or reached through incremental rise. When $U$ is raised in small steps of 0.2 V, no definitive patterned state appears below about 8 V, although director fluctuations increase with $U$, as evident from the increased contrast of the inhomogeneity as exemplified in (a) and (b); at 8 V, flexoelectric domains nucleate randomly and grow as in (c) showing an image from a time series. By contrast, soon after a sudden voltage of about 8 V is applied, ZZ domains possibly of electroconvective type form (d), but begin to decay in time (e, f); within a couple of seconds, flexoelectric domains nucleate (see encircled region in (e)) and slowly grow (f). Planar 5O9 sample at $T_r$= 2 ºC; P(5)–A(90); $d$=5 μm.

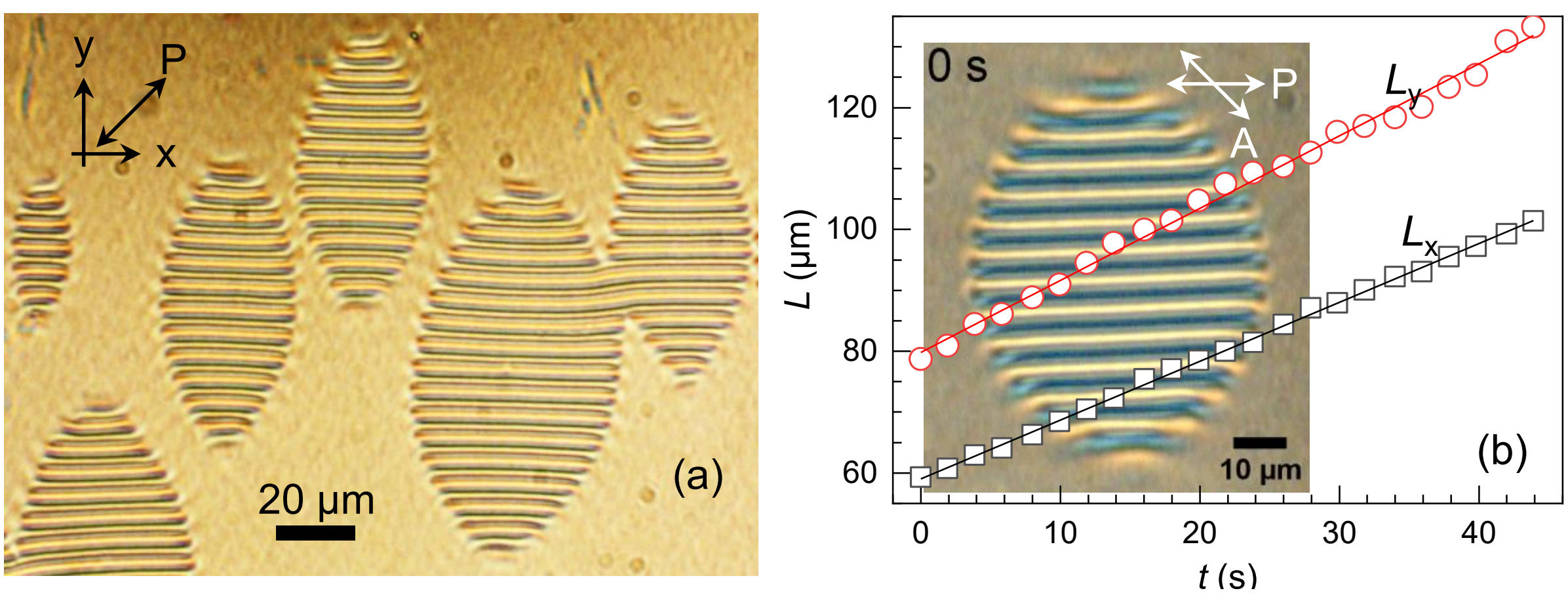

FIG. 7. Evolution of the periodic flexo-instability in a 5 μm thick nematic layer of 5O9 at $T_r$=2 ºC. (a) Typical tactoid-like geometry of localized flexoelectric islands nucleated randomly and developing outward at 7.5 V. (b) The linear-in-time variation of the principal dimensions $L_x$ and $L_y$ in a flexoelectric island (seen in the background) at 7.9 V.

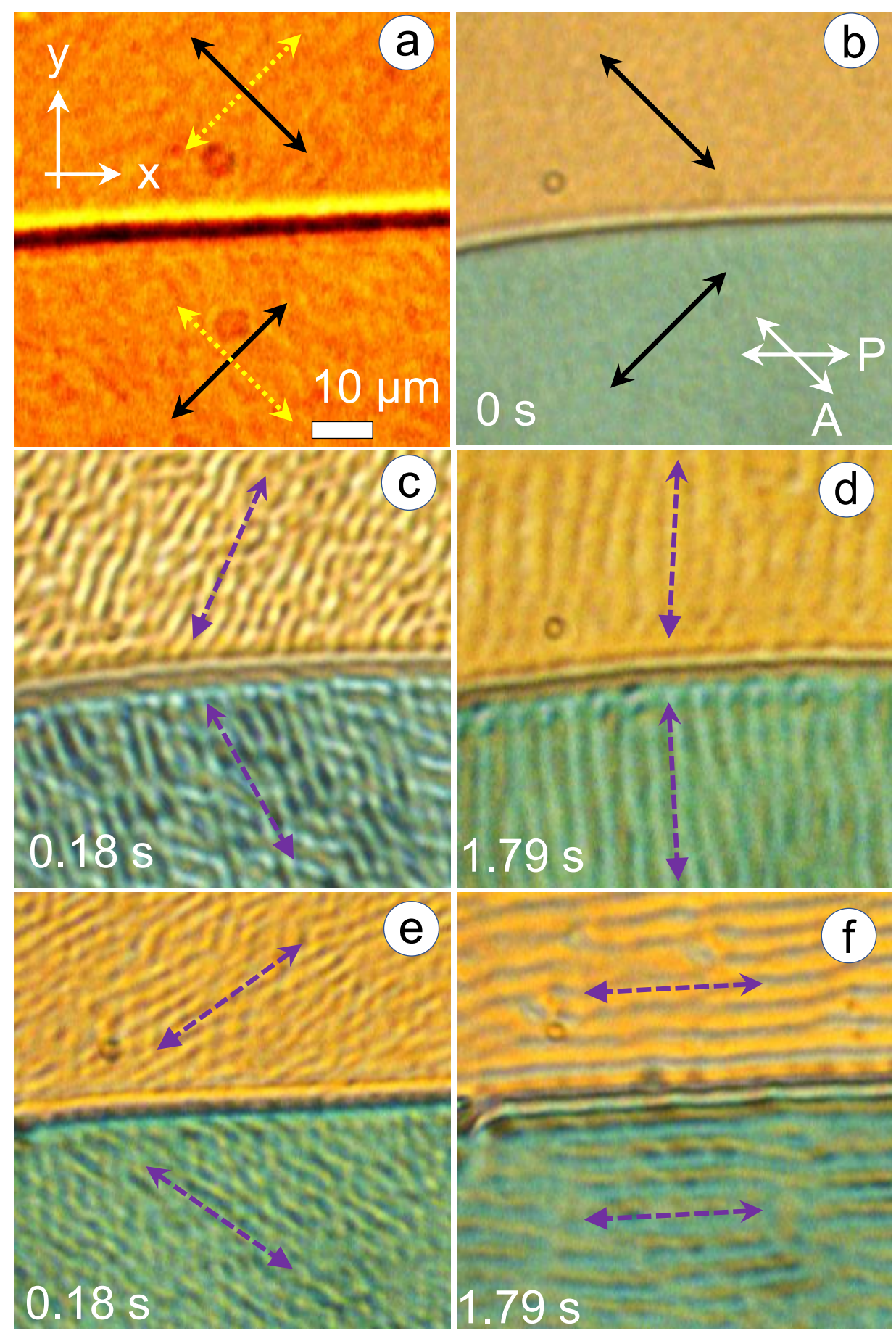

FIG. 8. Polarity sensitive, transient, EC instability localized near the positive electrode in a 90º twist cell containing a 5 µm thick nematic layer of 5O9 at $T_r$=2 ºC. (a) Field free ground state showing diagonal stripes in oppositely twisted separated by a near-horizontal disclination. Dotted (yellow) double arrows in the two regions indicate the stripe directions orthogonal to the midplane director (dark double arrows). (b-d) Select frames of a time series; a dc voltage of –7 V is imposed suddenly after frame (b) showing the base state at $t$=0 s; oblique stripes inclined more towards $y$ compared to $x$ form soon after the voltage is applied (c); with time, the stripes tend increasingly toward $y$ (d), even as they decay toward the base state. (e, f) Select frames of a time series recorded after a dc voltage of +7 V is applied at once; soon after imposing the field, oblique stripes inclined more towards $x$ compared to $y$ form (e); with time the stripes tend increasingly toward $x$ (g), even as they decay to result in the base state.

Fig. 7(b) illustrates the linear time dependence of principal axes of an enlarging flexo-instability region. The pattern within a flexo-island remains steady throughout, with the period unaffected by its size, while nucleation of new domains along $y$ and growth of existing domains along $x$ occur at constant rates. An interesting question concerns the location of maximum distortion amplitude in patterned states; it is expected to be the midplane region by symmetry, provided no surface polarization effects are involved. One simple way to ascertain the presence or otherwise of such effects is to examine the given instability using a 90º twist cell. We illustrate in Fig. 8 the results obtained in regard to the ZZ EC pattern formed with this geometry in nematic 5O9 subjected to a dc field. The figure shows two oppositely twisted regions separated by a near horizontal twist disclination of half-strength. In Fig. 8(a), the midplane director disposition is indicated relative to the stripes of the ground state; the top and bottom halves are respectively right and left-handed. Figs. 8(b)-8(d), and 8(e) and 8(g) show select frames of two time-lapse recordings made with $U$= –7 V and +7 V, respectively. The texture in Fig. 8(c) observed under a negative field, immediately after switch on, is that of oblique stripes inclined closer to $y$ compared to $x$. The inclination to $y$ reduces with time and a pattern of nearly vertical stripes of lowered contrast [Fig. 8(d)] is attained over a few seconds, before the pattern almost vanishes. A similar sequence of changes is observed with a positive field, except that the stripes then are more inclined towards $x$ at start and are nearly horizontal just before decaying completely. All these pattern characteristics show that the plane of maximum distortion, initially located close to midplane, drifts subsequently toward the positive electrode (anode). The progressive change of the wave vector direction toward that corresponding to the surface electroconvection state is explicable by considering the intrinsic surface electric field [39]. Assuming positive ions as selectively adsorbed (resulting in a surface charge density $\Sigma$), the intrinsic double-layer field will be along $+z$ at the bottom electrode and $-z$ at the top electrode; see Fig. 9. Soon after applying an external voltage $U$, the initial displacement profile (dashed curve) will be shifted by $D_e$ corresponding to $U$ (dotted line). Due to the overall increase in field, the EC instability occurs at first between the midplane and anode. As the equilibrium displacement profile (continuous line) is approached, the field at the anode increases even as the bulk field decreases so that the plane of maximum modulation drifts toward the anode.

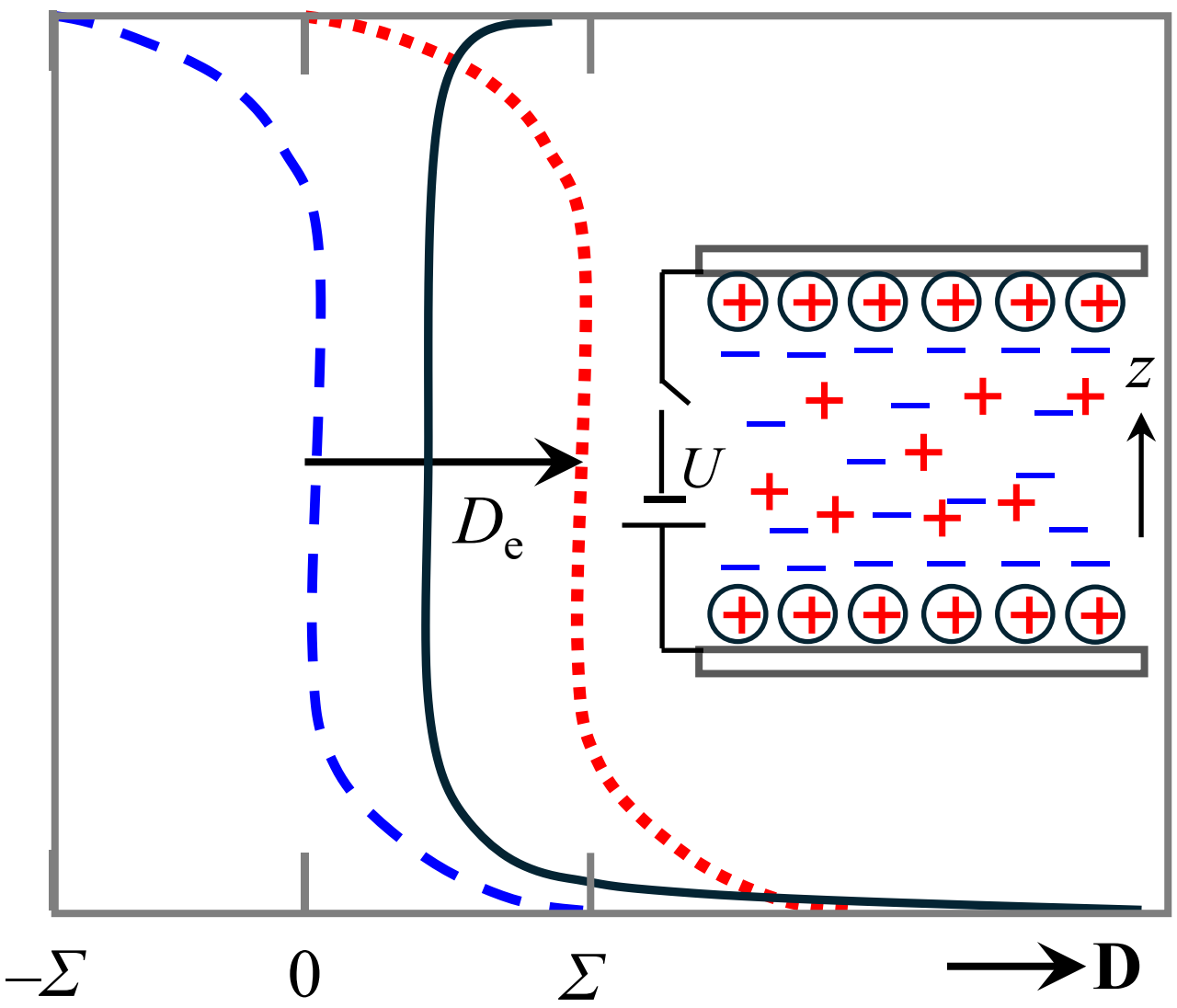

FIG. 9. Schematic of electric displacement profiles in the absence of applied voltage (dashed blue line), immediately after applying a voltage (dotted red line) and after ion redistribution under bias (continuous black line); $\Sigma$ denotes the density of charges adsorbed and $D_e$ the external-voltage-induced displacement along $z$. (inset) nematic layer with intrinsic double layers.

At higher voltages, i.e., above the threshold of Bobylev-Pikin flexo-instability, the surface EC stripes persist—albeit with diminished distortion amplitude—even at the nucleation of flexoelectric domains. This feature is exemplified in Fig. 10 showing the different pattern states in two oppositely twisted regions of 6O9 in a 90° twist cell; upon imposing +8.1 V suddenly, the first bifurcation into the EC state leads to a pattern of horizontal (h-) stripes in *both* regions; this instability is clearly localized near the anode (bottom electrode, Fig. 2); also seen in Fig. 10(a) showing the h-stripes are flexodomains nucleated within the encircled regions; as the latter form in bulk, they stretch nearly along the midplane director determined by the twist sense. The h-stripe state is progressively eliminated by enlarging flexodomains [Fig. 10(b). A sudden imposition of –8.1 V leads initially to vertical (v-) stripes of EC at the top electrode (anode) and flexo instability replaces EC eventually, as before [Fig. 10(c)]. As the voltage is increased above the Bobylev-Pikin instability threshold, the morphological features of the flexo-pattern undergo a series of striking changes. Of these, the well-established feature is the variable grating effect, namely the strong voltage dependence of the domain width $w_F$ given by

$$w_F = \frac{\pi k d}{1.2 e U}, \qquad (8)$$

where $e = e_1$ (splay flexo-coefficient)= – $e_3$ (bend flexocoefficient), $k_{11}$ (splay elastic modulus)= $k_{22}$ (twist elasic modulus)=$k$ and permittivity anisotropy $\varepsilon_a$ is assumed 0; this equation is predicted [40] to hold good qualitatively even if $e_1 \neq - e_3$ and $\varepsilon_a \neq 0$, provided $|\mu| = |\varepsilon_o \varepsilon_a k / (e_1 - e_3)^2| < 1$, so that the domain density $L = 1/w_F$ is linear in $U$. In the two dimers studied here, we do find L to increase linearly with voltage (Fig. 11).

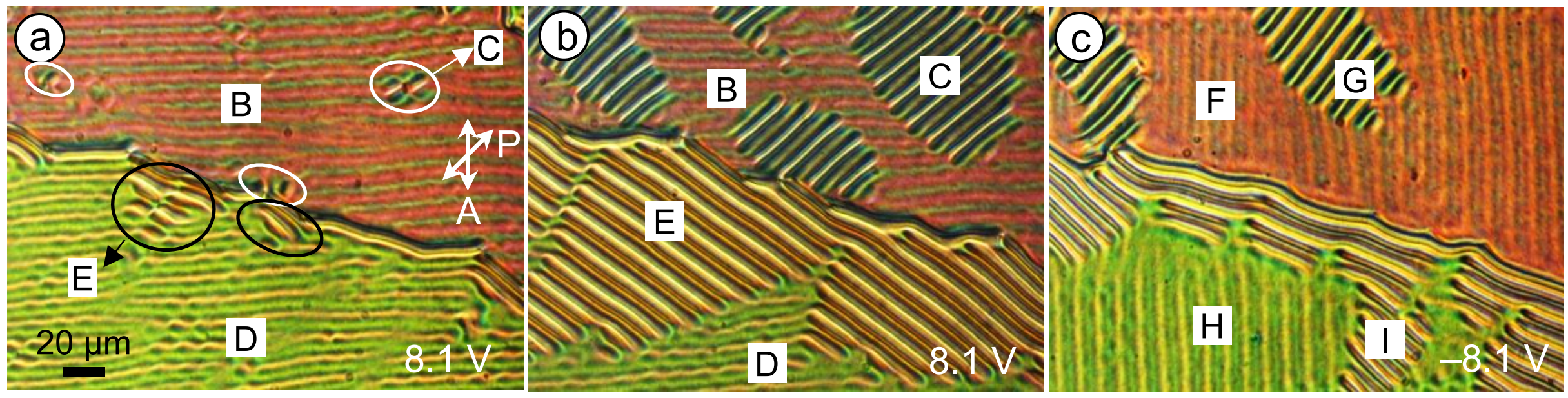

FIG. 10. Instabilities excited by a static electric field in a 90° twisted planar 10 μm-thick layer of 6O9 at $T_r$=3.5 °C. (a) With a positive voltage of 8.1 V, primary bifurcation leads to horizontal (h-) stripes in both the reverse twisted regions (labelled B and D) close to the onset of flexo-instability; these stripes due to electroconvection, are localized in the vicinity of the bottom electrode (anode). In time, secondary bifurcation due to the volume flexoelectric instability appears in the form of nearly diagonal stripes within sporadically nucleated encircled regions (labelled C and E). (b) Flexoregions grow at the expense of EC rolls; flexostripes, which deviate slightly from the midplane director, occur close to, but not along, the opposite diagonals in the two twisted regions. (c) On reversing the polarity, at –8.1 V, the primary EC instability appears as vertical (v-) stripes (regions F and H) again at the anode (top electrode); flexodomains (regions G and I) gradually replace the EC stripes.

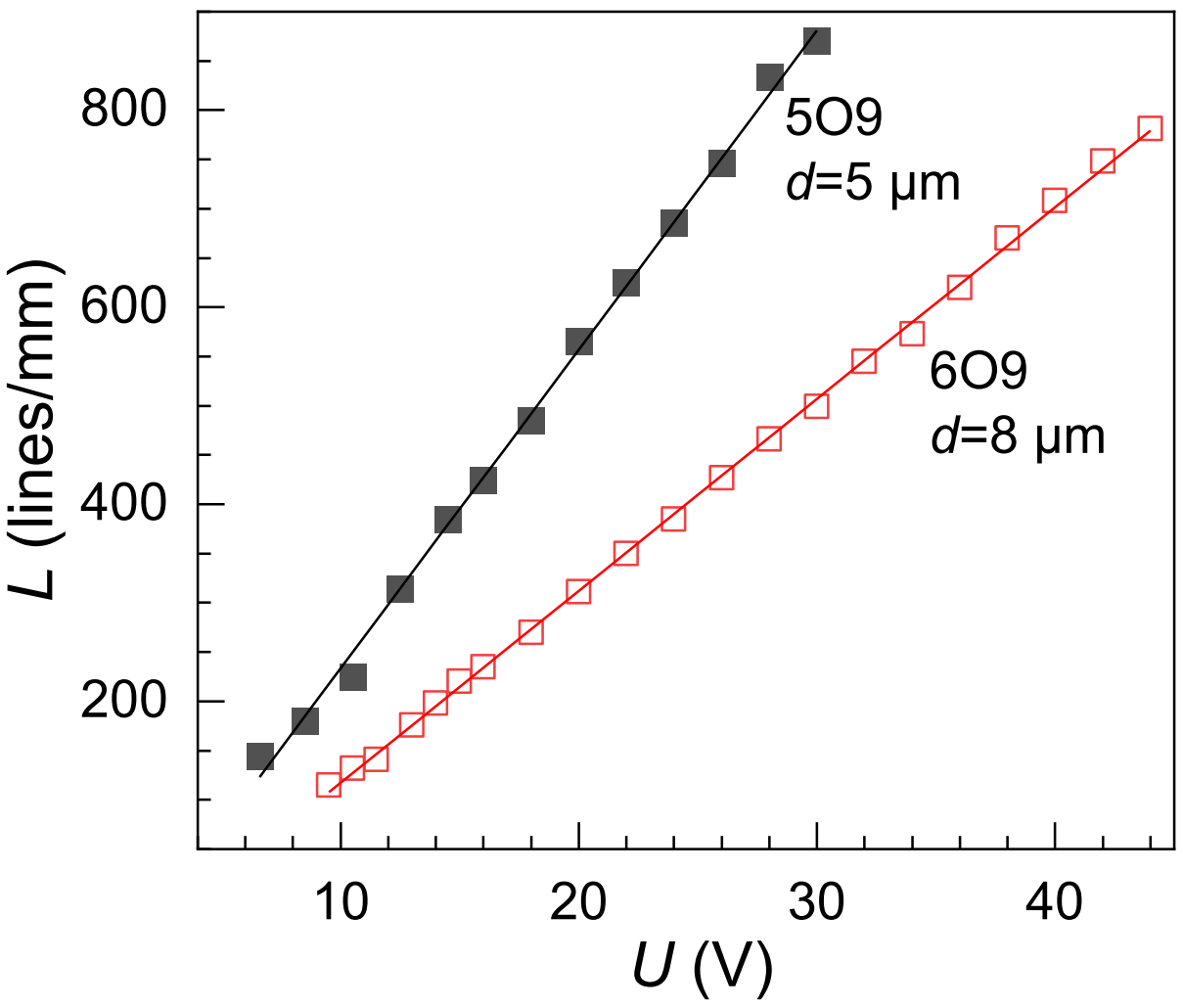

FIG. 11. Line density of flexoelectric domains as a function of applied voltage in nematic layers of 5O9 and 6O9 at $T_r$=6.8 ºC and 3.5 ºC, respectively.

As $w_F$ decreases, half-strength disclinations of opposite topological charge are generated increasingly. Unlike in calamitic nematics, but much as in rigid bent-core nematics [41], the wavevector orientation in the layer plane becomes increasingly random with rising voltage, leading to the formation of dipole and quadrupole geometries of topological defects [Fig. 12(a)-12(c)]. The optical pattern progressively assumes the appearance of the fingerprint texture in cholesterics; eventually, when the domain density becomes too large for the stripes to be seen apart, the texture appears to be made of focalconic-fan-like objects, as in Fig. 12(d). This resemblance of the unresolved periodic flexoelectric state to the smectic state of a layered lattice has been attributed, in the case of rigid bent-core nematics, to the bend type distortion that is energetically favoured in bent-core systems compared to splay, unlike in calamitics. This reasoning is likely to be applicable more to nematics made of flexible-core bent molecules compared to rigid-core bent molecules in view of the relatively negligible bend elastic modulus in the former.

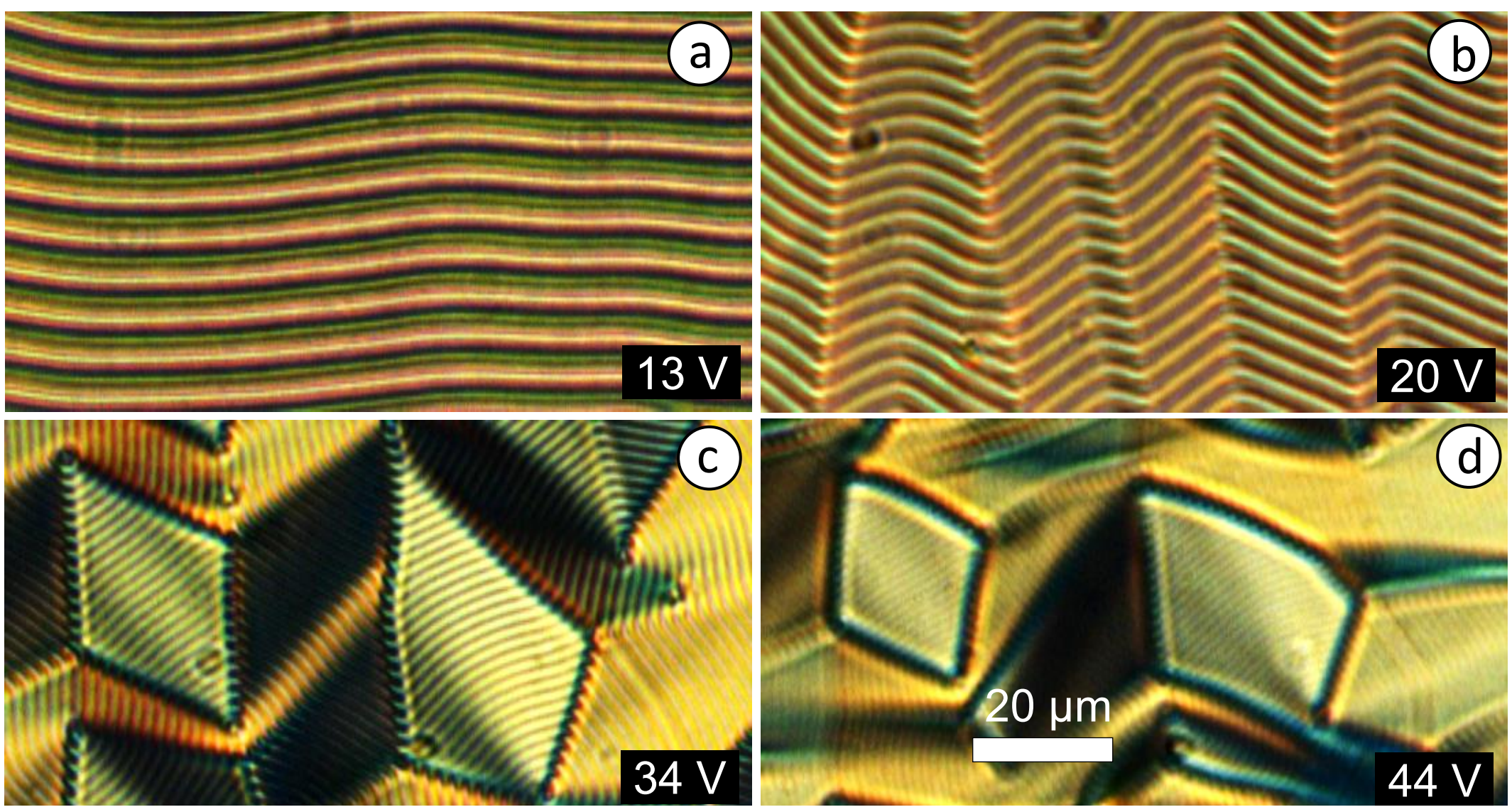

FIG. 12. Evolution of focalconic-fan-like texture in a planar 8 μm thick layer of 6O9 with increasing voltage; $T_r$=3.5 °C. The domain lines nearly along $x$ in (a) turn zizag in (b); in (c), the entire region is filled with angular quadrupoles (fans), and the domains within them are still well resolved; in (d), the domains are beginning to be hazy and unresolved.

**D. Electroconvective and flexoelectric states under very low-frequency (<1 Hz) ac fields**

The periodic states excited by very low frequency electric fields are also illustrative of the competing electrohydrodynamic and flexoelectric instabilities. The alignment modulations that follow the application of a sinewave field across a nematic layer are of the type seen in Fig. 13; it depicts some select frames $F_{ii}$ of a time image series from a 6.8 μm thick untwisted planar layer of nematic 6O9, at $T_r$=3.5 ºC, exposed to a 12.6 V sine wave field of frequency 0.05 Hz. Frames $F_1$ and $F_{73}$ show the field-free dark state at $U_t$=0 V. The first or low-voltage instability appears as oblique rolls of EC ($F_8$), which develops into the bimodal EC state ($F_{25}$) at a higher $U_t$; flexodomains nucleate randomly within the bimodal state and extend all over ($F_{57}$) before the sample returns to the base state ($F_{73}$) at the next polarity reversal. Similar modulations occur during the following half-cycle ($F_{73}$-$F_{130}$). In the schematic, Fig. 13(a), the time-voltage positions for the textures in various frames are indicated. Fig. 13(b) shows the profiles of the imposed voltage $U_t$ (V) and associated optical response $R_o$. The first $R_o$ peak is due to the formation of oblique stripes. Even as they begin to decay, increasing $U_t$ revives and amplifies the instability into the bimodal EC state resulting in the second $R_o$ peak. Within the bimodal state, flexodomains originate locally and grow. The third $R_o$ peak is entirely due to flexoelectric domains well defined under P(0)-A(90) and disappearing soon after. Between $F_{25}$ and $F_{57}$, i.e., around the peak voltage, flexodomains of large distortion amplitude display varied and vivid birefringence colours. After the polarity switch, after 10 s, the same sequence of textural modifications follows. From Fig. 13 referring to the instabilities in a planar cell, localization of EC near the electrode surfaces is not inferable. However, experiments using the 90º-twisted planar geometry and very low frequency sine wave or square wave fields, clearly demonstrate the appearance of EC at the surfaces.

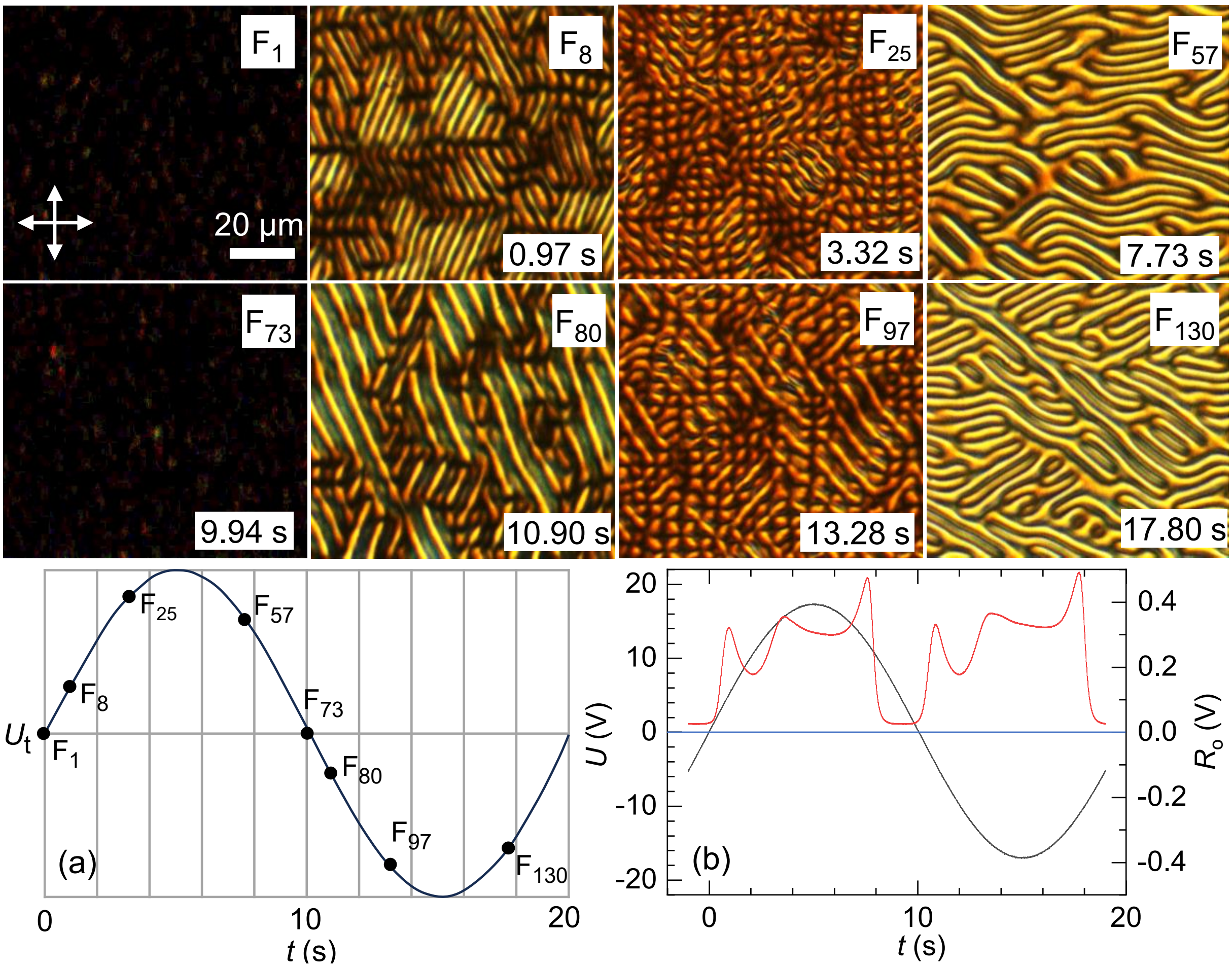

FIG. 13. Typical sequence of periodic modulations observed in very low frequency fields exemplified with select frames of time image series from a 6.8 μm thick planar layer of nematic 6O9 at $T_r$=3.5 ºC exposed to a 12.6 V sine wave field of frequency 0.05 Hz. $F_{ii}$ are the frame numbers and successive frames are 0.1383 s apart. The ($U_t$, $t$) positions corresponding to different frames are indicated by dots in the schematic (a). $F_1$ and $F_{73}$ show the field free dark state at $U_t$=0 V. The primary instability seen as oblique rolls is EC ($F_8$), which develops into the bimodal EC state ($F_{25}$) with increasing $U_t$; flexoinstability nucleates within the bimodal and extends all over ($F_{57}$) before the sample returns to the base state ($F_{73}$). The same sequence of modulations occurs during the following half-cycle ($F_{73}$-$F_{130}$) with minor morphological changes. The profiles in (b) show the imposed voltage $U_t$ (V) and associated optical response $R_o$ (measured in voltage corresponding to the transmitted light intensity, using a photodiode coupled to a transimpedance amplifier).

Transient EC and flexoelectric periodic states that occur separated in time within each half-cycle of a driving low-frequency (<100 mHz) electric field have previously been observed in the dielectrically negative rodlike nematogens Merck N4 [42], Merck Phase V [43] and 4-n-octyloxyphenyl 4-n- methyloxybenzoate [44]. The flashing character of both dissipative and equilibrium instabilities are interpreted in terms of the extended standard model of EC that takes into account the flexoelectric influence, but not double layer effects [43, 44]. EC flashes, as in the present study, precede flexo-flashes; they are phase locked to current spikes forming soon after voltage reversals at times noted as dependent on ionic reorganization. These earlier observations are on planar untwisted samples of Merck Phase V [43] and a rodlike phenyl benzoate [44]. Our finding of EC as a polarity sensitive *surface* effect shows the importance of selective adsorption of ions and the need to consider it in any analysis of time varying field across the sample. We may mention, in passing, that the flexoelectric pattern existing at any static voltage $U_{DC}$ is completely suppressible by simultaneous application of a high frequency ac voltage $U_{AC}$.

### E. Electroconvection in low frequency (5 Hz-10 kHz) fields

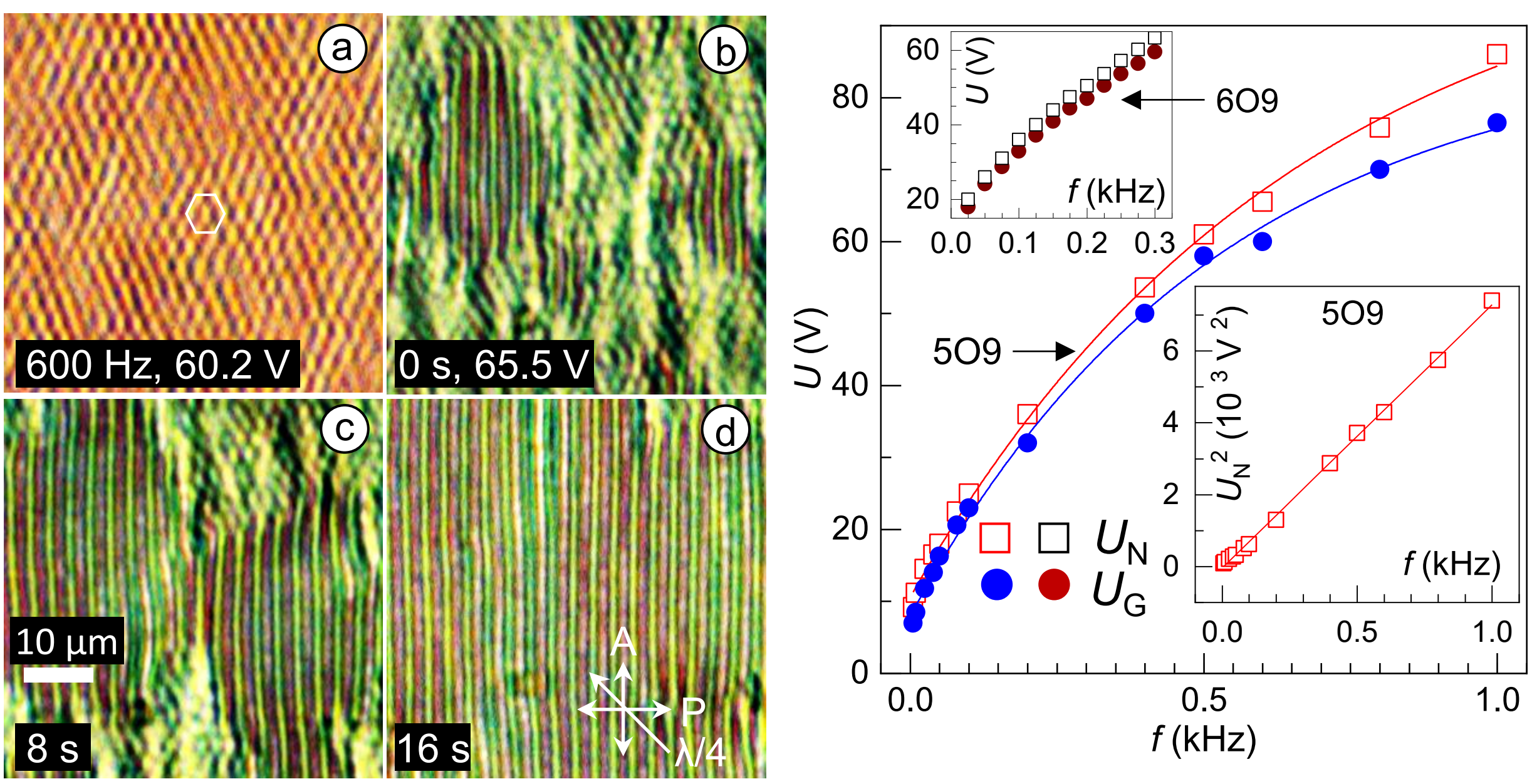

FIG. 14. (Left) Discontinuous transition from the regular zz state (grid state) of oblique rolls to the normal roll state, observed in a 5 µm thick planar layer of 5O9 at $T_r$=2 °C. (a) The 2-D grid texture prior to the transition; 600 Hz (square

wave, SQW), 60.2 V. (b-d) Select frames of a time series recorded at 65.5 V showing the sporadic islands of normal stripes that grow with time.

FIG. 15. (Right) Parabolic increase of the threshold voltages $U_G$ and $U_N$ at which bifurcations take place, respectively, into the grid and normal-roll states. Inset of 5O9 shows the linear increase of $U_N^2$ with $f$.

In the mHz regime, as seen above, the Bobylev-Pikin flexoelectric modulation completely suppresses electroconvection at a sufficiently high voltage. Above ~1 Hz, its exclusive appearance ceases; flexopolarization may, however, enter electrohydrodynamical equations to cause the wave vector of convective rolls to deviate from the direction of $\mathbf{n}_o$. In the frequency region of 5 Hz-1 kHz, the first bifurcation in both the dimers is to zigzag rolls. That these rolls form in bulk is clear from their diagonal orientation in 90°-twisted planar samples of either dimer. Often, slightly above threshold, the oblique rolls superimpose so regularly as to generate a hexagonal network of focal images (grid state) with a repeat distance of ~2 μm (along $x$) in 5 μm thick samples. This feature, depicted in Fig. 14(a), is best visualized using a diagonal polarizer P(45), or crossed polarizers P(45)-A(135) with a quarter wave plate. As the voltage is raised progressively, the grid state transitions discontinuously, at a well-marked voltage $U_{PC}$, into what may be called the pre-chevron (PCh) state; the latter manifests in normal stripes with the wave vector along the initial director $\mathbf{n}_o$ and a spatial period of ~3 μm in layers of 5 μm thickness; Figs. 14(b)-14(d) illustrate the growth of sporadically formed domains of this secondary PCh instability. Characteristically, the PCh stripes display varied colours over a range of focal distances. The thresholds of the two bifurcations increase parabolically in both 5O9 and 6O9, as exemplified in Fig. 15. During the growth of the PCh state, the contiguous oblique rolls reduce their inclination relative to the $y$-axis. Further, the PCh state may be induced to undergo the reverse transition by decreasing $U$ to less than $U_{PC}$. Close examination of the stripes in the normal roll state reveals them to be spotty due to a quasiperiodic intensity variation along their length. In fact, on elevating $U$ marginally above $U_N$, the spots expand into inclined loops with opposite edge dislocations. This is an unusual course of the third-stage development leading to a chevron state. As it happens, in 5O9, the fast dynamics of chevrons, particularly the climbing motion of dislocations, does not allow for a uniform texture to establish at any time [Fig. 16(a)]; in 6O9, however, a slower dynamic situation allows for a regular pattern to prevail over longer periods [Fig. 16(b)]. With a progressive increase in $f$, the formation of the chevron state occurs at a correspondingly increasing $U=U_{CH}$, with $U_{CH}^2$ *vs* $f$ being nearly linear. The variation of $U_{CH}$ with $f$ follows the same trend as in Fig. 15 above, with $U_{CH}$ exceeding 120 V at 10 kHz. To prevent sample degradation under high electric fields, we did not examine the frequency region between 10 kHz and 100 kHz. In the following subsection, we deal with a nonstandard instability observed above 100 kHz, at voltages below 100 V.

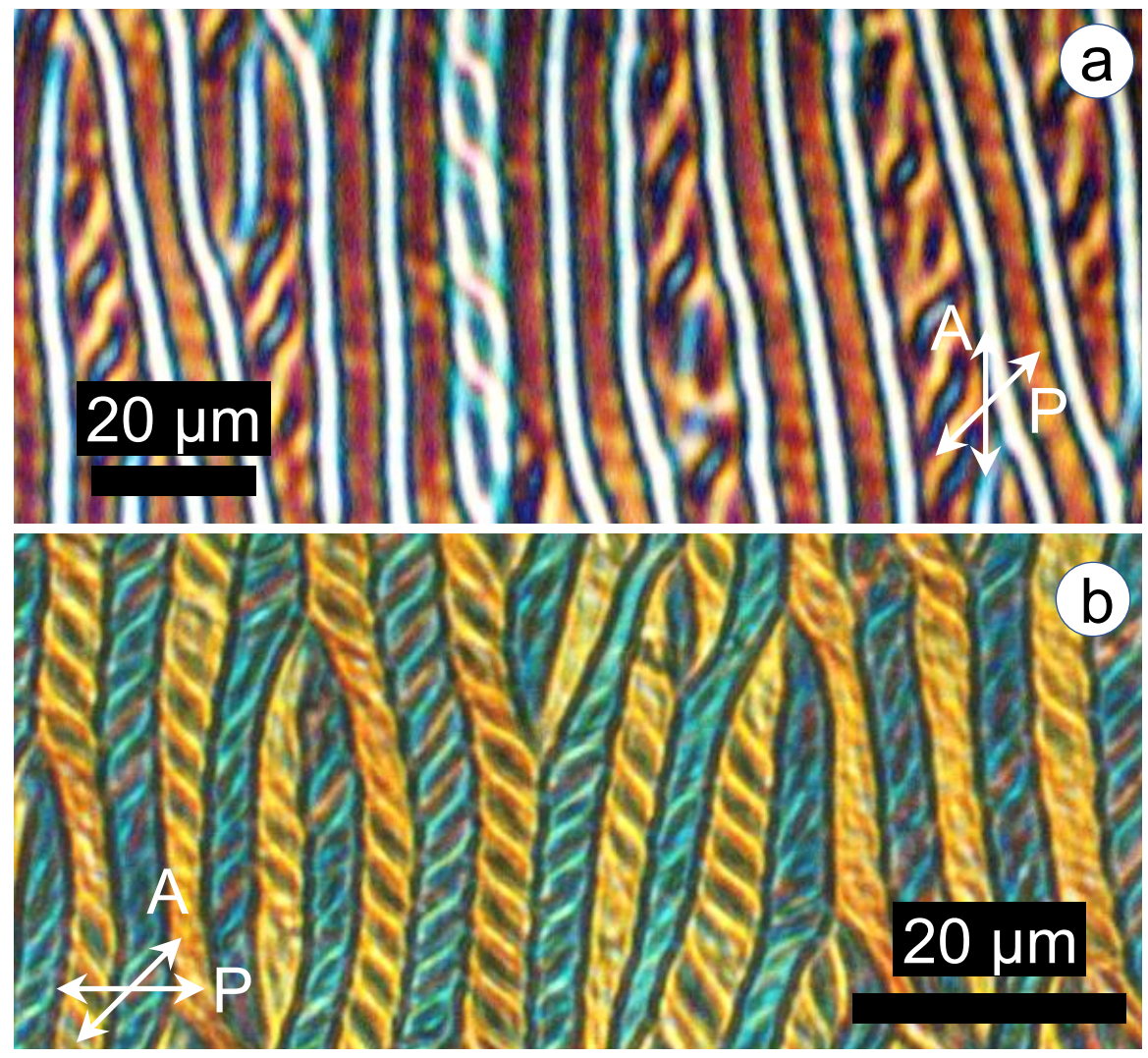

Fig. 16. The chevron state excited by SQW fields in (a) 5O9 ($T_r$=2 ºC, $f$=25 Hz, $U$=14.8 V) and (b) 6O9 ($T_r$=3.5 ºC, $f$=20 Hz, $U$=19.5 V). The patterns vary spatiotemporally with faster dynamics in 5O9 compared to 6O9.

### F. Very high frequency (≥100 kHz) instabilities

As noted in the introductory section, bent rigid-core nematics (BRCNs) exhibit two distinctive modes of EC appearing optically as wide normal rolls (NR or INR to signify associated inplane flows), predominantly due to periodic azimuthal distortion [34]. Of these, the mode belonging, relatively, to the lower frequency regime occurs at a voltage threshold $U_{NR}$ that increases nonlinearly with frequency. By contrast, $U_{NR}(f)$ is a decreasing function for the far-frequency mode. In the bent flexible-core nematics (BFCNs) under study, this latter mode is clearly observed. As the voltage is increased, at first, the ground state narrow stripes gain in amplitude to appear with ever increasing contrast. The NRs begin to develop as a superimposed structure over the ground state at a sufficiently high voltage $U_{NR}$. In Figs. 17(a) and 17(b), the narrow and wide stripes appearing in the hybrid state are shown separated in terms of their focal planes. Between partially crossed polarizers and at a higher voltage, the wide stripes appear vividly, as in Fig. 17(c).

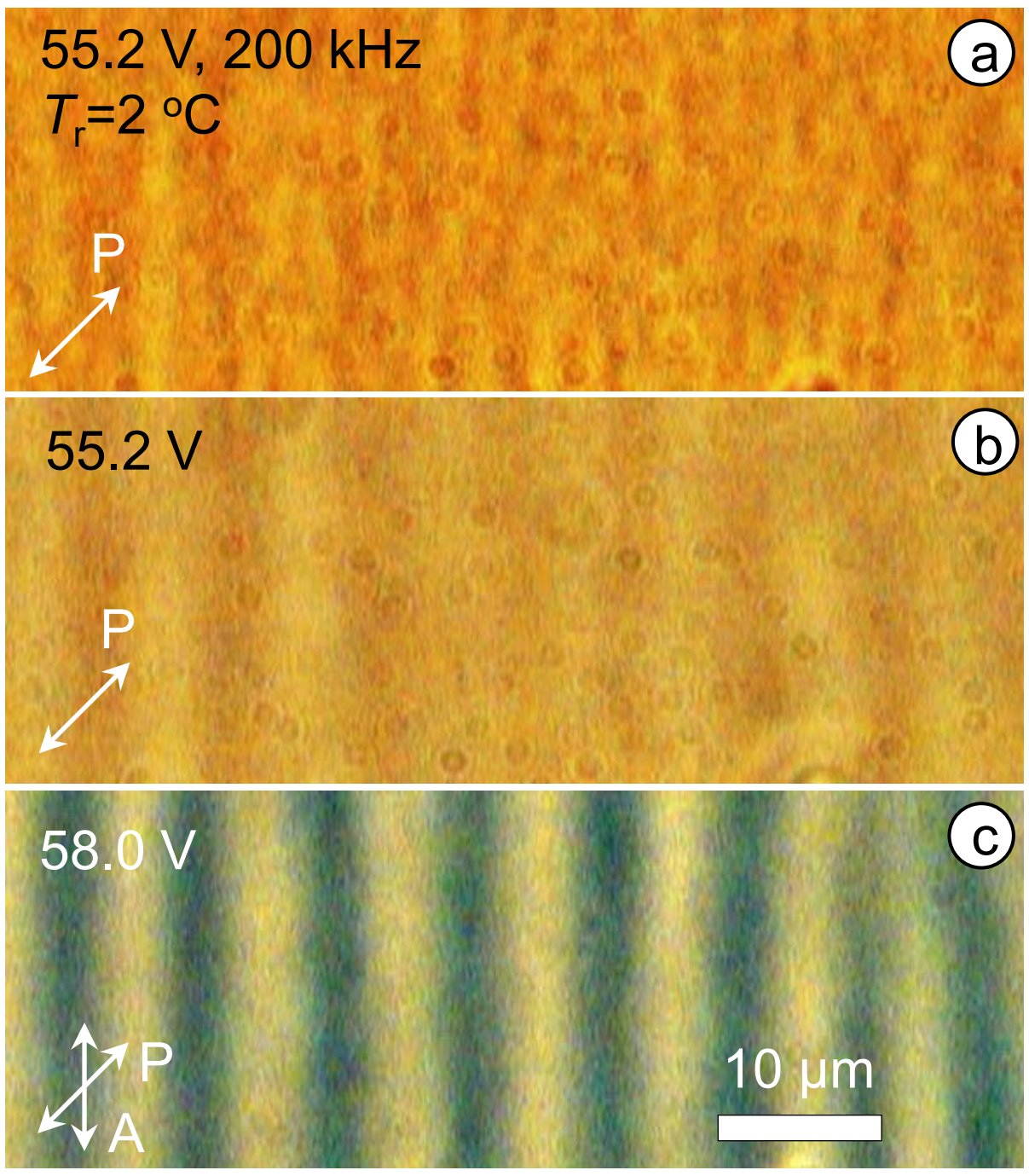

FIG. 17. High frequency, wide domains evolving superimposed over the ground state stripe structure in a 5 μm thick nematic layer of 5O9 at $T_r$=2 °C. The focus in (a) is on the quasiperiodic ground state, and that in (b) is on wide domains; the two focal planes are about 20 μm apart. (c) Wide domains seen clearly at a higher voltage, between partially crossed polarizers.

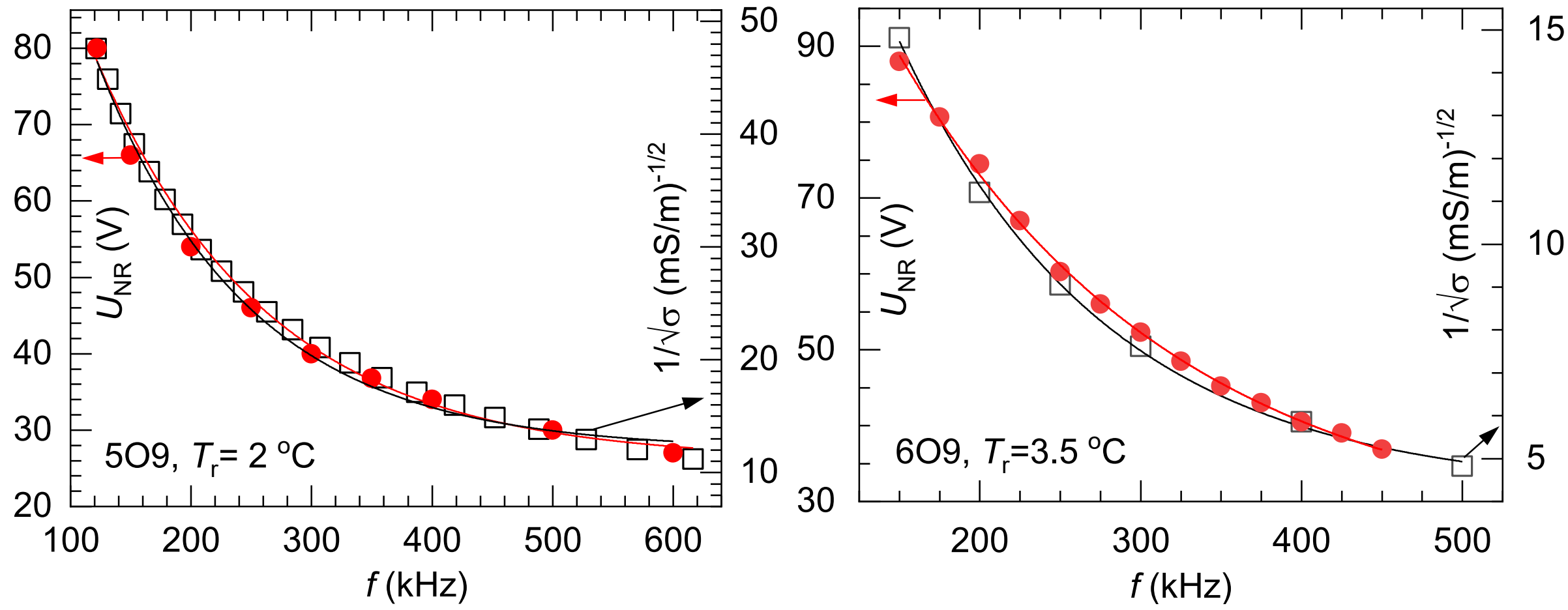

FIG. 18. Similar exponential frequency variations of $U_{NR}$ (the threshold of wide NRs) and $1/\sqrt{\sigma}$ ($\sigma = \sigma_{\perp}$ being the electrical conductivity along the director-normal) in the relaxation region of $\varepsilon_{\perp}$in (a) 5O9 and (b) 6O9. Dark circles and red squares correspond, respectively, to $U_{NR}$ and $1/\sqrt{\sigma}$. Notable is the similar scaling behavior of $U_{NR}$ and $1/\sqrt{\sigma}$ with respect to $f$ in the two dimers.

The voltage threshold $U_{NR}$ decreases exponentially with increasing frequency; a similar variation is also found with $1/\sqrt{\sigma_{\perp}}$, where $\sigma_{\perp}$ is the electrical conductivity along the director-normal. These dependencies, common to both the dimers, are depicted in Figs. 18(a) and 18(b). We may note here that domains of the very high frequency regime are not unique to bent-core systems. Indeed, they were observed very early in calamitic systems [45,46]. However, the continued drop in $U_{NR}$ under increasing $f$ is a feature observed so far exclusively in bent-core nematics. Wide-domains in calamitics have been interpreted by Pikin and Chigrinov [40, 46] as due to the inertial term, usually ignored in electrohydrodynamical equations, becoming significant at very high frequencies. According to their analysis that takes this term into account, the far frequency regime is similar to the usual *conduction* regime except for the volume charges oscillating with opposite phase relative to the field, and the director pattern and flow velocity varying weakly relative to their mean values; the usual dielectric regime is predicted to lie between the low-frequency and high-frequency conduction regimes. The threshold voltage of the inertial conduction instability is shown to be given, in SI units, by

$$U_{NR} = 2\pi f \sqrt{[K\, \varepsilon\, \varepsilon_o / \{\sigma\, (\sigma_{\parallel} - \sigma_{\perp})\}]}, \qquad (9)$$

where K, $\varepsilon$ and $\sigma$ are the effective values of elastic constant, dielectric permittivity and electrical conductivity, respectively; and ($\sigma_{||}- \sigma_{\perp}$) is the conductivity anisotropy. The two dimers of this study behave very similarly in respect of frequency variation of $\varepsilon_{\perp}$ and $\sigma_{\perp}$. As seen in Fig. 19, the relaxation occurs over a wide range, starting around 100 kHz and extending beyond 1 MHz. It is in this region that the conductivity rises sharply (see insets in Fig. 19). We were unable to determine $\varepsilon_{||}$ and $\sigma_{||}$ since homeotropic alignment could be obtained in commercial cells with ITO plates treated for such an alignment. However, during a voltage ramp-up, with increasing amplitude of out-of-plane distortion (associated with either the quasiperiodic ground state or a field-induced periodic order), the effective permittivity falls continuously from $\varepsilon_{\perp}$, testifying to the dimers being dielectrically negative; this is demonstrated in the $\varepsilon'$-$U$ plot. In further confirmation of the sign of $\varepsilon_a$, the birefringence colour remains unaffected even at very high fields provided no field-induced periodic instability exists at the chosen $f$ ; for example, a planar 7 μm thick layer of 5O9 at $T_r$=2 °C, shows a steady cyan hue even up to 100 V at 50 kHz.

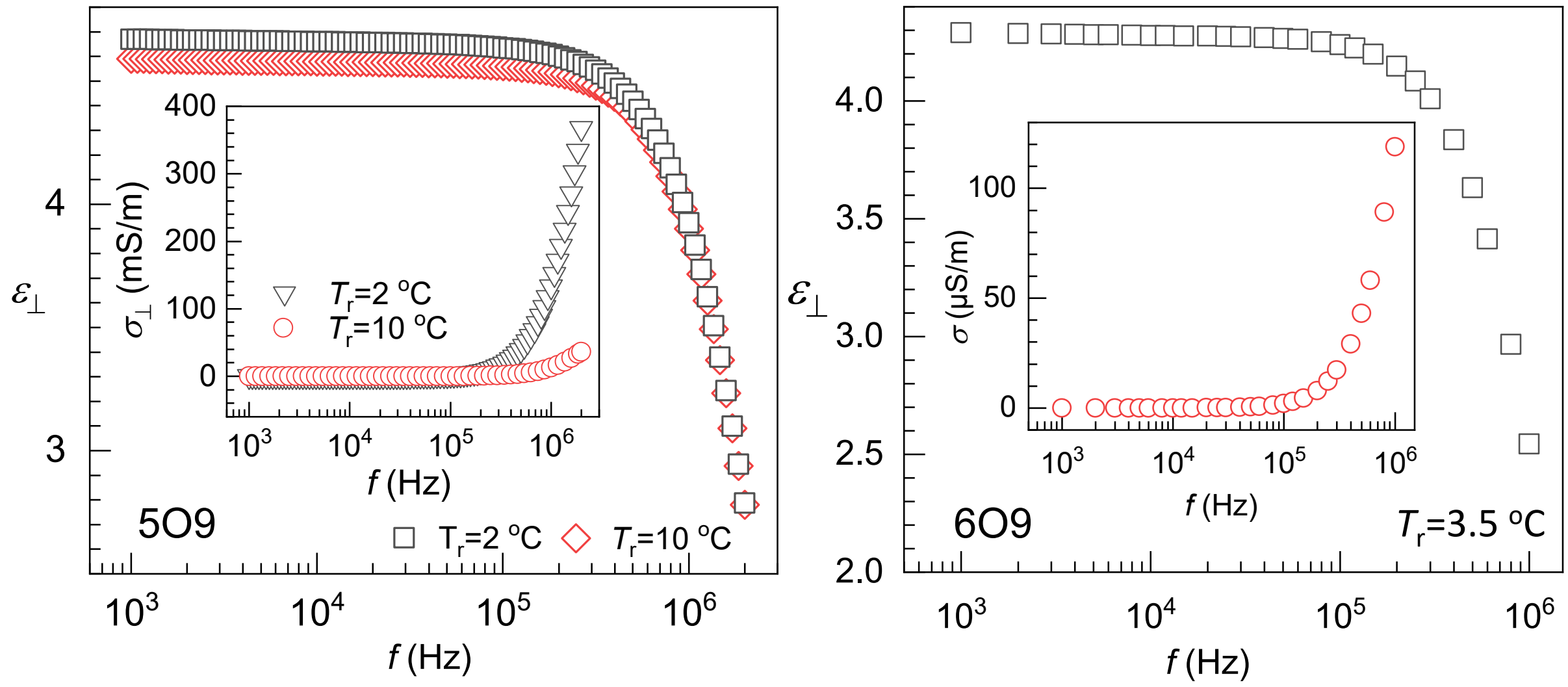

FIG. 19. Frequency variation of dielectric permittivity and electrical conductivity, measured along the director normal, in 5O9 and 6O9.

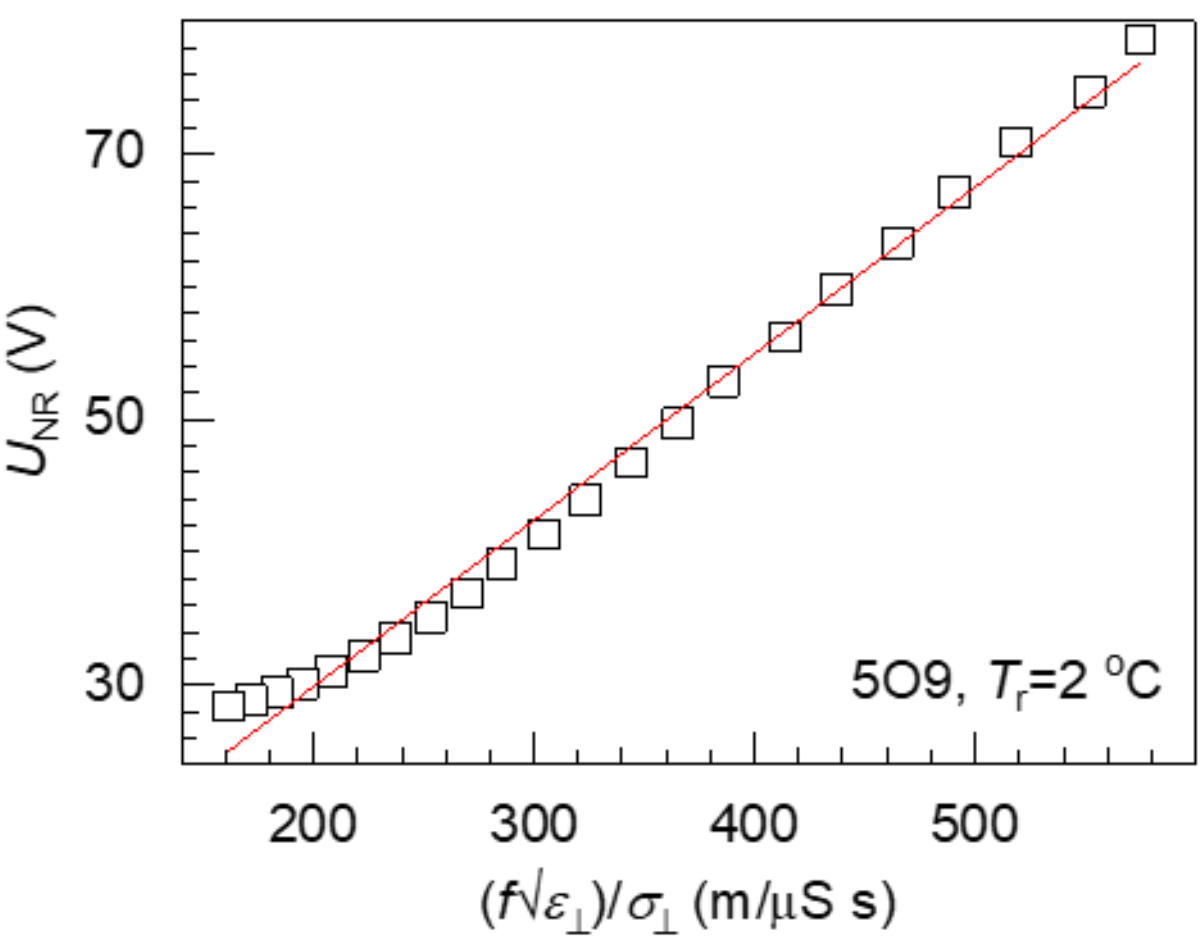

FIG. 20. Near linear variation of $U_{NR}$ with $f\sqrt{\varepsilon_{\perp}}/\sigma_{\perp}$; data are from a 5 μm thick planar N layer of 5O9 at $T_r$=2 °C.

In order to broadly test the applicability of Eq. (9) to our results, we assume $\sigma_{\perp}$ to represent both $\sigma$ and ($\sigma_{||}- \sigma_{\perp}$), and $\varepsilon$=$\varepsilon_{\perp}$; we then find that, in both 5O9 and 6O9, $U_{NR}$ is nearly linear with respect to $f\sqrt{\varepsilon_{\perp}}/\sigma_{\perp}$. From the slope of the fit-line, 0.125 V nS s/m, we obtain the effective elastic constant

K as ~1 pN, which is of the correct order considering the various assumptions. It seems, therefore, reasonable to identify the far frequency "nonstandard" domains as due to the inertial conduction instability; notably it develops in the region of relaxation of $\varepsilon_\perp$. In fact, Pikin and Chigrinov did envisage the possibility of $U_{NR}(f)$ being a decreasing function in the dielectric relaxation region, noting that "the product $\sigma(\sigma_\| - \sigma_\perp)$… depends strongly on the frequency $\omega$, increasing with increasing $\omega$, and by the same token weakens strongly the frequency dependence of the threshold voltage; in particular, it is possible in principle that the threshold voltage decreases with increasing frequency".

The period $\lambda_{NR}$ of the INR pattern in 5O9 and 6O9 is not very sensitive to frequency and remains at approximately $2d$ over a wide range of frequencies. This feature is similar to the earlier finding in rigid bent-core nematics [34]. With respect to voltage also, as observed in 6O9 for voltages in the range 70-89 V at 150 kHz, $\lambda_{NR}$ does not show a significant dependence, remaining equal to about $2d$. While the pattern contrast increases with $U$ at a given $f$, or with $f$ at a given $U$, no other change is observed. Due to the instrumental limitation on maximum $U$ obtainable at high frequencies, any secondary instabilities beyond the INR could not be explored. Moreover, the quantitative results obtained using very high $U$ and $f$ values are to be treated with caution as trend-indicators in view of the heating effects expected.

## IV. CONCLUSIONS

We have presented experimental studies on the patterned states in two flexible bent-core dimers with a negative dielectric anisotropy, in their nematic phase that overlies the $N_{TB}$ phase. It is a generic feature in bent-core nematics that their ground state is inhomogeneous; this manifests optically in fluctuating quasiperiodic stripes with the wave vector along the alignment direction. We argue that this instability has its origin in the competition between flexoelectric and elastic energies of the azimuthally quasiperiodic base structure. The rest of the report concerns various electrically induced periodic states; the observations covered in this part are conveniently represented in the schematic phase diagram; see Fig. 21. Static and quasistatic fields excite competing surface electroconvective and volume flexoelectric modes, manifesting in oblique roll (OR) and flexodomain (FD) patterns, respectively. Uncommonly, the flexodomains appear localized in tactoid-like regions [Fig. 7(a)]. There are, indeed, several studies, both theoretical and experimental, on nematic 3-dimensional, spindle-like tactoids [47]; however, there does not seem to be any analysis of 2-dimensional tactoids in the electrically excited patterned state. Thus (as suggested by one of the Referees), it would be an interesting problem for future analysis to examine the role of applied electric field in relation to the geometry of the confining region within which periodic structures are nucleated.

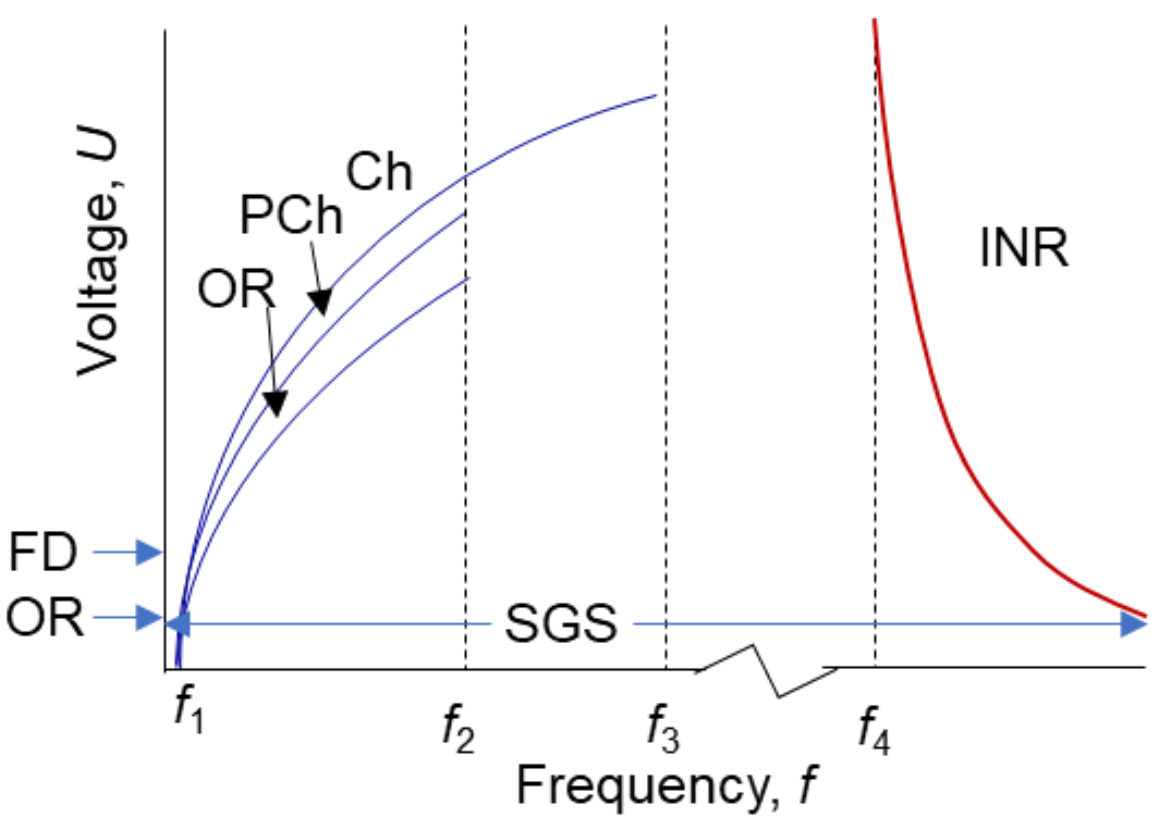

FIG. 21. Schematic phase diagram of the dimers 5O9 and 6O9 in the *U-f* plane. Quasiperiodic *striped ground state* (SGS) underlies all field-induced states. In DC and very low frequency (mHz) fields, *oblique rolls* (OR) of electroconvection appearing as a surface instability give way, at higher voltages, to *flexoelectric* domains (FD). Between $f_1$ and $f_2$, the sequence of states under increasing $U$ is OR →*Pre-Chevron*, PCh →*Chevron*, Ch. Between $f_2$ and $f_3$, only the Ch state is definitively discerned as developing from the SGS. The interval $f_3$-$f_4$ is not explored to keep $U$<120 V to minimize any electrochemical degradation. Beyond $f_4$ nonstandard *inplane normal rolls*, INR, appear that show a continued lowering of threshold voltage with increasing $f$ in conformity with the inertial mode of electroconvection excited under large values of $\omega\varepsilon''$.

Between $f_1$ and $f_2$ (5 Hz and 1 kHz in 6O9), under increasing $U$, EC states corresponding to oblique roll, pre-chevron and chevron patterns are obtained, in that order. Above $f_2$, only the chevron state is definitively discerned as developing from the SGS, with the transition threshold being an increasing function of $f$; the upper frequency limit of this region lies above $f_3$ (10 kHz in 6O9) and remains undetermined due to the voltage limitation. INR state of the flexible dimers belonging to the very high frequency regime (above 100 kHz and extending to the MHz region in both dimers) is similar to that in rigid bent-core nematics. Interestingly, it conforms to the predictions of the Pikin-Chigrinov theory that takes the inertia term into account.

## ACKNOWLEDGMENTS

The authors are thankful to Prof. B. L. V. Prasad, Director, Centre for Nano and Soft Matter Sciences, Bangalore for the experimental facilities.